\documentclass[aps,floatfix,nofootinbib, twocolumn,superscriptaddress,amsmath,amssymb,amsfonts]{revtex4-1} 

\usepackage[T1]{fontenc}
\usepackage[latin9]{inputenc}
\usepackage{lmodern} 
\usepackage{color}
\usepackage{bbold}
\usepackage{dsfont}
\usepackage{graphicx}
\usepackage[caption=false]{subfig}
\usepackage[colorlinks]{hyperref}
\usepackage[bottom]{footmisc}
\usepackage{enumerate}
\usepackage{float}
\usepackage{mathtools}

\usepackage{empheq}
\usepackage{tikz}
\usepackage{fancyhdr}
\usepackage[normalem]{ulem}
\DeclareMathAlphabet\mbc{OMS}{cmsy}{b}{n}
\usepackage{balance}

\begin{document}

\global\long\def\eqn#1{\begin{align}#1\end{align}}
\global\long\def\vec#1{\overrightarrow{#1}}
\global\long\def\ket#1{\left|#1\right\rangle }
\global\long\def\bra#1{\left\langle #1\right|}
\global\long\def\bkt#1{\left(#1\right)}
\global\long\def\sbkt#1{\left[#1\right]}
\global\long\def\cbkt#1{\left\{#1\right\}}
\global\long\def\abs#1{\left\vert#1\right\vert}
\global\long\def\cev#1{\overleftarrow{#1}}
\global\long\def\der#1#2{\frac{{d}#1}{{d}#2}}
\global\long\def\pard#1#2{\frac{{\partial}#1}{{\partial}#2}}
\global\long\def\re{\mathrm{Re}}
\global\long\def\im{\mathrm{Im}}
\global\long\def\dd{\mathrm{d}}
\global\long\def\ddd{\mathcal{D}}

\global\long\def\avg#1{\left\langle #1 \right\rangle}
\global\long\def\mr#1{\mathrm{#1}}
\global\long\def\mb#1{{\mathbf #1}}
\global\long\def\mc#1{\mathcal{#1}}
\global\long\def\tr{\mathrm{Tr}}
\global\long\def\dbar#1{\Bar{\Bar{#1}}}

\global\long\def\nth{$n^{\mathrm{th}}$\,}
\global\long\def\mth{$m^{\mathrm{th}}$\,}
\global\long\def\non{\nonumber}

\newcommand{\orange}[1]{{\color{orange} {#1}}}
\newcommand{\cyan}[1]{{\color{cyan} {#1}}}
\newcommand{\blue}[1]{{\color{blue} {#1}}}
\newcommand{\yellow}[1]{{\color{yellow} {#1}}}
\newcommand{\green}[1]{{\color{green} {#1}}}
\newcommand{\red}[1]{{\color{red} {#1}}}
\global\long\def\todo#1{\orange{{$\bigstar$ \cyan{\bf\sc #1}}$\bigstar$} }

\title{Quantum Brownian Motion of a particle from Casimir-Polder Interactions}

\author{Kanupriya Sinha}
\email{kanu@umd.edu}
\affiliation{US Army Research Laboratory, Adelphi, Maryland 20783, USA}
\affiliation{Joint Quantum Institute, University of Maryland, College Park, MD 20742, USA}
\author{Yi\u{g}it Suba\c{s}\i}
\email{ysubasi@lanl.gov}
\affiliation{Computer, Computational and Statistical Sciences Division, Los Alamos National Laboratory, Los Alamos, NM 87545, USA}
\begin{abstract}

We study the fluctuation-induced dissipative dynamics of the quantized center of mass motion of a polarizable dielectric particle trapped near a surface. The particle's center of mass is treated as an open quantum system coupled to the electromagnetic field acting as its environment, with the resulting  system dynamics described by a quantum Brownian motion master equation. The dissipation and decoherence of the particle's center of mass are characterized by the modified spectral density of the electromagnetic field that depends on surface losses and the strength of the classical trap field.  Our results are relevant to experiments with levitated dielectric particles near surfaces, illustrating potential ways of mitigating fluctuation-induced decoherence while preparing such systems in macroscopic quantum states.
\end{abstract}

\maketitle

\section{Introduction}
Creating macroscopic superpositions of massive systems as a means to understand the quantum-to-classical transition is a task of foundational importance \cite{Zurek91}. Among  promising experimental platforms for realizing large superpositions of massive objects, levitated optomechanical systems bring together the advantages of optical trapping and cooling methods in terms of control, while being well-isolated from an environment in the absence of mechanical clamping, thus minimizing decoherence \cite{Vamivakas16, Yin, Oriol11, OriolPRA11, Bateman14}. There has been astonishing experimental  progress in terms of control and manipulation of levitated dielectric nanoparticles -- ranging from recent demonstrations of  cooling particles down to micro- and milli-kelvins, \cite{Uros19, Carlos19, Novotny19}, to observation of rotational frequencies as large as  MHz-GHz with remarkable stabilities \cite{Ahn18, Reimann18, Kuhn17}.

Interfacing such precisely controlled mesoscopic quantum systems with waveguides further allows for better manipulation and probing mechanisms of the system of interest, as guided photonic modes can couple efficiently to particles in the near-field regime \cite{Diehl18, Thomson13, Juan16}.  Near-field levitated nanophotonics can therefore allow for a strong optomechanical couplings of mesoscopic systems  with well-controlled fields, as has been demonstrated in \cite{Magrini18}.

 However, when preparing a system  in a macroscopic quantum state near surfaces, one needs to consider that the quantum (and thermal) fluctuations of the electromagnetic (EM) field are enhanced due to the presence of the surface degrees of freedom \cite{Volokitin07}. The increased density of EM field modes can therefore cause the system of interest to decohere faster in the vicinity of a surface, as has been shown both theoretically and experimentally with regard to the internal degrees of freedom of particles near surfaces \cite{ spinflip1, spinflip2, superchip, Hinds1999, Schumm2005, Lin2004,  Jones2003}. It is similarly imperative to analyze the fluctuation-induced decoherence for the external degrees of freedom in near-field nanomechanical experiments  \cite{Skatepark}.

In this paper  we  study the decoherence and dissipation of the quantized center of mass (COM) motion of a neutral dielectric particle trapped near a surface. We show that the   open system dynamics of the particle can be described in terms  of the  quantum Brownian motion (QBM) master equation \cite{HPZ92, BPBook,Lombardo04}, and the surface-modified dissipation and decoherence can be expressed in terms of a modified spectral density of the electromagnetic field. We further draw a correspondence between the surface-induced decoherence of the particle's COM  and the collisional model of decoherence \cite{JoosZeh, MaxBook}.

The paper is organized as follows. In section \ref{model} we develop a theoretical model of a polarizable dielectric particle interacting with  the EM field in the presence of a surface, deriving the  fluctuation- and drive-induced potentials. In section \ref{results} we derive the QBM master equation for the quantized COM motion, and analyze the resulting decoherence and dissipation of the particle for different surface properties in section \ref{dissdec}. We summarize our findings in section \ref{discussion}.

\section{Model}
\label{model}

\begin{figure*}[ht]
    \centering
\subfloat[]{ \includegraphics[width = 3 in]{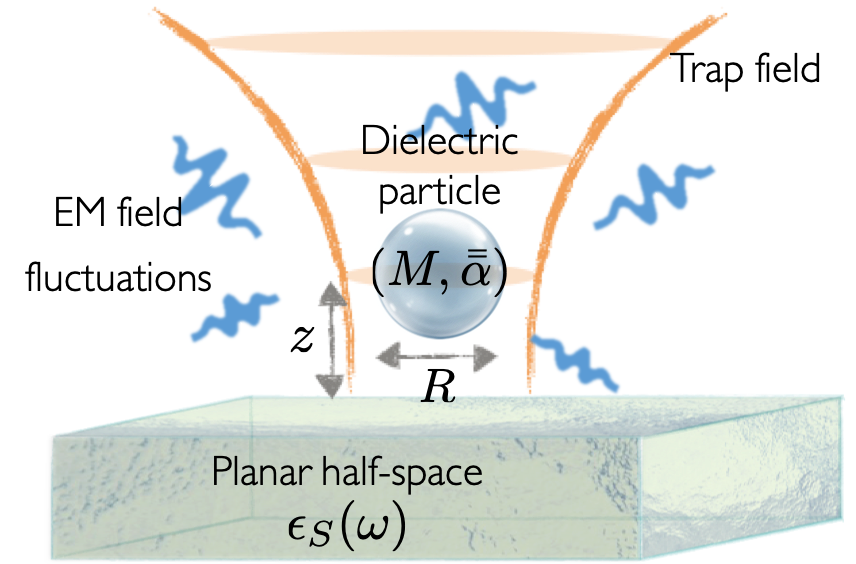}}
\subfloat[]{    \includegraphics[width = 3 in]{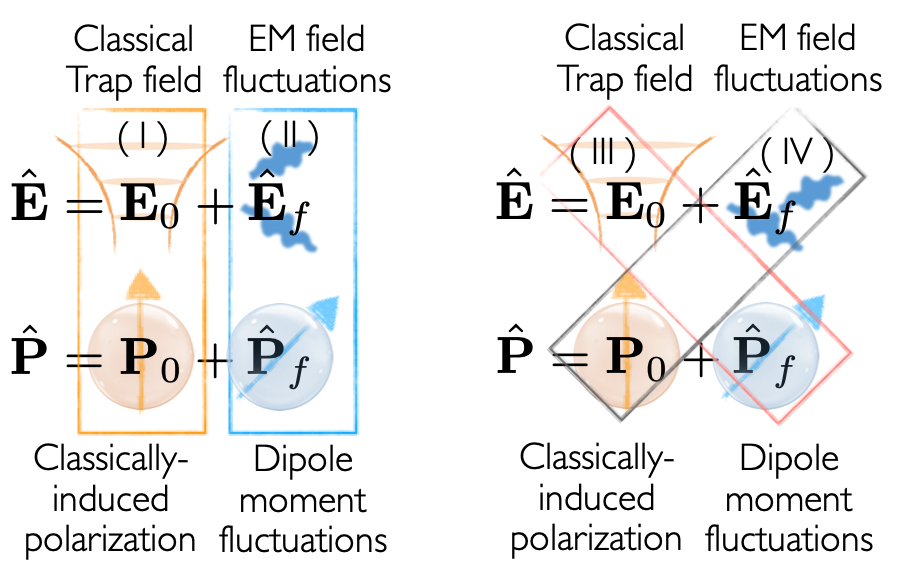}}
    \caption{(a) Schematic representation of a dielectric particle of mass $M$, radius $R$ and polarizability $\dbar{\alpha}\bkt{\omega}$ near a planar half-space with permittivity $\epsilon_S\bkt{\omega}$, interacting with the quantum fluctuations of the EM field. We assume a classical trap field incident normally  on the surface which creates a standing wave potential perpendicular to the surface. The particle is trapped in the first intensity maxima of the standing wave potential, at a distance $z$ from the surface. (b) The four contributions to the total interaction energy as a result of the interaction between the total electric field and the induced polarization of the particle -- (I) corresponds to the classical trap potential $\hat H_\mr{Tr}$ due to the interaction of the trap field with the clasically induced polarization (Eq.\,\eqref{htr}). (II) stands for the Casimir-Polder interaction $\hat H_\mr{CP}$ arising from the interaction between the fluctuations of the EM field with the fluctuating dipole moment of the particle (Eq.\,\eqref{hcp}). (III) and (IV) taken together lead to the drive-induced Casimir-Polder Hamiltonian $\hat H_\mr{DCP}$ (Eq.\,\eqref{hdcp}). (III)  arises from the interaction between the classical field and the dipole moment fluctuations and (IV) corresponds to the interaction between the classically induced polarization and the EM field fluctuations. }
    \label{schematic}
\end{figure*}

Let us consider a  dielectric particle of mass $M$ and polarizability $\dbar{\alpha}\bkt{\omega}$  placed at a distance $z$ from a planar half-space medium of permittivity $\epsilon_S \bkt{\omega}$, as depicted in Fig.\,\ref{schematic}\,(a).  We consider a classical driving field that is incident normally on the surface of the medium and reflected to form a standing wave potential. We assume that the particle is trapped near the first intensity maxima of the standing wave potential near the surface \cite{Magrini18}.

The Hamiltonian for the total system can be written as\eqn{\hat H =\frac{\hat{p}_z^2}{2m} + \hat H_F + \hat H_\mr{int},}where the first term corresponds to the kinetic energy of the particle with  $ \hat{p}_z$ as the quantized COM  momentum  along the $z$-axis. In the absence of a strong transverse confinement we ignore the quantized motion in the $xy$-plane.  $\hat H_F$ corresponds to the Hamiltonian of the quantized field in the presence of the medium (see Eq.\,\eqref{hv}). The interaction Hamiltonian $\hat H_\mr{int}$ represents the electric-dipole interaction between the  polarizable particle and the total electric field over the volume $V$ of the particle given by \cite{Jackson}
\eqn{\label{hint0}\hat H_{\mr {int}} = - \int_V \dd^3 r' \hat {\mb{P}}\bkt{{\mb{r}}'} \cdot \hat{\mb {E}}({\mb{r}}' ),}
where $\hat {\mb{P}}\bkt{\mb{r}'}$ refers to the polarization of the dielectric, $\hat{\mb {E}}(\mb{r}')$ is the electric field, with $\mb{r}'$ being a point in  the volume of the particle. 

The total electric field at a position $\mb{r}$ can be expressed as $\hat{\mb {E}}\bkt{\mb{r}'} = {\mb {E}}_0\bkt{\mb{r}' ,t} +  \hat{\mb {E}}_{f}\bkt{\mb{r}' } $, where ${\mb {E}}_0\bkt{\mb{r} '} $ is the classical trap field, and $\hat{\mb {E}}_f\bkt{\mb{r} '}  $ refers to the fluctuations of the field (quantum and thermal). We assume that the trap field is given as
\eqn{\label{E0}
\mb{E}_0 \bkt{{\mb{r}'}, t} = \frac{1}{2}\sbkt{{ \mbc{E}}_0\bkt{{\mb{r}'}} e^{-i \omega_0 t}+{ \mbc{E}}_0^\ast\bkt{{\mb{r}'}} e^{i \omega_0 t}} ,
}
with $\omega_0 $ as the frequency, and $\mbc{E}_0\bkt{\mb{r}'}$ as the amplitude of the electric field at position $\mb{r}'$ which takes the incident and reflected fields into consideration. We further assume that the field is polarized along the $xy$-plane.

Using the macroscopic QED formalism  \cite{SYB1, SYB2}, the electric field fluctuations in the presence of a surface are given as (see \eqref{Era})
\eqn{\label{Ef}
&\hat{\mb{E}}_f\bkt{\mb{r}'} =\non\\ &\int\dd\omega\sum_{ \lambda = e,m}\int \dd^3 r
\sbkt {\dbar {G}_\lambda \bkt{\mb{r}', \mb{r}, \omega}\cdot \hat{\mb{f}}_{\lambda}\bkt{\mb{r},  \omega} + \text{H.c.}},
}
where $\dbar{G}_\lambda \bkt{\mb{r}_1 , \mb{r}_2, \omega}$ stands for the propagator of a field excitation between points $\mb{r}_1 $ and $\mb{r}_2$, as described by Eq.\,\eqref{Ge}--\eqref{Gtot} \cite{GreenWelsch}.

Assuming that the particle has a linear, homogeneous and isotropic polarizability $\dbar\alpha\bkt{\mb{r}',\omega} \equiv \alpha \bkt{\omega} \mathbb{1}$, we can write the induced polarization  of the particle as $\hat{\mb{P}}\bkt{\mb{r}'} ={\mb{P}}_0 \bkt{\mb{r}',t}+ \hat{\mb{P}}_f\bkt{\mb{r}'}$ \cite{Jackson}, where
\eqn{\label{P0}{\mb{P}}_0 \bkt{\mb{r}',t}= \frac{1}{2}\sbkt{\alpha\bkt{\omega_0 } \mbc{E}_0\bkt{\mb{r}'} e^{-i \omega_0 t}+ H.c.}
}
is the polarization induced by the classical field and
\eqn{\label{Pf}
 &\hat{\mb{P}}_f\bkt{\mb{r}'} =\non\\ &\int\dd\omega\,\sum_{ \lambda = e,m}\int \dd^3 r\sbkt{
 \alpha\bkt{\omega }\dbar {G}_\lambda \bkt{\mb{r}', \mb{r}, \omega}\cdot \hat{\mb{f}}_{\lambda}\bkt{\mb{r},  \omega} + \text{H.c.}},
}
 corresponds to the polarization induced due to the fluctuations of the EM field. 
 
 We now assume that the dielectric particle is point-like, with the COM position of the particle $\hat{\mb{r}}_M =\hat{\mathbb{1}} \mb{r}+\hat{z}\mb{e}_z$  such that $\mb{r}$ corresponds to the classical COM coordinates and $\hat z$ represents the quantum fluctuations of the COM motion along the $z$-axis about the classical trap position.

Using Eq.\,\eqref{E0}--\eqref{Pf}, one can rewrite the interaction Hamiltonian Eq.\,\eqref{hint0} as \eqn{\hat H_{\mr{int}} =  \hat H_{\mr{Tr}}+ \hat H_\mr{CP} + \hat H_\mr{DCP} ,}  where \eqn{\label{htr}\hat H_\mr{Tr}\equiv -{\mb{P}}_0 \bkt{\hat{\mb{r}}_M,t } \cdot \mb{E}_0 \bkt{\hat{\mb{r}}_M, t} ,
} corresponds to the trap Hamiltonian,  \eqn{\label{hcp}\hat H_\mr{CP}\equiv -{\hat{\mb{P}}}_f \bkt{\hat{\mb{r}}_M} \cdot \hat{\mb{E}}_f \bkt{\hat{\mb{r}}_M} ,} stands for the Casimir-Polder (CP) interaction Hamiltonian, and
\eqn{\label{hdcp}\hat{H}_{\mr{DCP}}&=\non\\
&-\sbkt{ {\mb{E}}_0 \bkt{\hat{\mb{r}}_M, t} \cdot \hat{\mb{P}}_f \bkt{\hat{\mb{r}}_M}+ \hat{\mb{E}}_f \bkt{\hat{\mb{r}}_M} \cdot {\mb{P}}_0 \bkt{\hat{\mb{r}}_M, t} },}
is the driven Casimir-Polder (DCP) interaction. The first term in the above stands for the fluctuating dipole interacting with the classical trap field, and the second term corresponds to  the classically driven dipole interacting with the EM field fluctuations at the position of the particle, as depicted by the processes (III) and (IV) in Fig.\,\ref{schematic}\,(b). We  study each of these contributions in detail in the following.
\subsection{Classical Trap}
The classical trap potential to zeroth order in the COM fluctuations is given as \eqn{\label{utr}U_\mr{Tr} \bkt{\mb{r}}\equiv -\frac{1}{2} \avg{\mb{P}_0 \bkt{\mb{r}, t}\cdot \mb{E}_0 \bkt{\mb{r}, t}} = - \frac{1}{4}\alpha\bkt{\omega_0 }\abs{\mbc{E}_0\bkt{\mb{r}}}^2,
}
where we have taken a time-average over the electric field. We note that the factor of $1/2$ is introduced to avoid the double sum of the energy associated with the interaction of the induced polarization and electric field \cite{Jackson}.  We have  further assumed here that the dielectric particle has negligible internal loss with a real  polarizability such that $\alpha\bkt{\omega} = \alpha^\ast\bkt{\omega}$.

Expanding $\hat H_\mr{Tr}$ to second order in the COM fluctuations $\hat z$ around the classical equilibrium position $\mb{r}_0$, and ignoring constant energy shifts,  one obtains the trap potential as
$\hat {V}_\mr{Tr} \equiv  \frac{1}{2}M \Omega_\mr{Tr}^2 \hat{z}^2,$
where $\Omega_\mr{Tr} = \sqrt{\frac{ \alpha \bkt{\omega_0 } k_0^2\abs{\mbc{E}_0 \bkt{\mb{r}_0 }}^2}{2M}}$ corresponds to the frequency of the trap due to the classical field. We have assumed here that the electric field amplitude for the standing wave formed by the classical trap field goes as $\mbc{E}_0\bkt{\mb {r}_0 } \sim e^{i k_0 z}$ as a function of $z$.
\subsection{Casimir-Polder Interaction}
Considering the interaction between field and polarization fluctuations  to zeroth order in COM fluctuations $\hat z$, such that $\hat H_\mr{CP}^{(0)}\equiv  - \hat{\mb{P}}_f \bkt{\mb{r}}\cdot \hat{\mb{E}}_f \bkt{\mb{r}}$, one can obtain the  Casimir-Polder potential as  $U_\mr{CP }(\mb{r}) \equiv \frac{1}{2}\tr_F \sbkt{ \hat\rho_F \hat {H}_\mr{CP}^{(0)}}$.  This can be evaluated in first order perturbation theory as \cite{SYB2}
\eqn{\label{ucp}
&U_{\mr{CP}} (\mb{r})  = \frac{\hbar\mu_0  }{2\pi }\int_0 ^\infty \dd\xi \xi ^2  \alpha\bkt{i \xi}\tr \sbkt{\dbar{G}_{\mr{sc}} \bkt{\mb{r}, \mb{r}, i\xi }} \non\\
&-\frac{\hbar\mu_0 }{\pi} \int \dd\omega \,\omega^2 n_\mr{th}\bkt{\omega} \alpha\bkt{\omega} \tr\sbkt{\im \dbar{G}_{\mr{sc}} \bkt{\mb{r}, \mb{r}, \omega}} ,
}
where we have assumed that the  field density matrix $\hat{\rho}_F$ corresponds to a thermal state   with temperature $T$ and  $n_\mr{th}\bkt{\omega} = \avg{\hat{\mb{f}}_\lambda^\dagger\bkt{\mb{r},\omega}\cdot\hat{\mb{f}}_\lambda\bkt{\mb{r}, \omega}}= \frac{1}{e^{\hbar \omega/(k_B T) }- 1}$ is the average number of thermal photons in the mode $\omega$.  All surface properties enter into consideration through the scattering Green's tensor $\dbar{G}_{\text{sc}}\bkt{{\bf r}, {\bf r}, i\xi}$ (see Eq.\eqref{Gtot}) corresponding to the  propagation of a virtual photon from the position $\bkt{{\bf r}}$ of the particle  to the surface and back. Given that imaginary frequencies are associated with virtual interactions, the above potential can be physically understood as coming from the interaction between the fluctuations of the dipole and those of the vacuum EM field, summed over all frequencies of virtual photons exchanged between the particle and the surface. The second term corresponds to the scattering and reabsorption of thermal fluctuations of the EM field by the particle off the surface.

\subsection{Drive-induced Casimir-Polder Interaction}
The linearized part of the interaction Hamiltonian with respect to the classical field gives a drive-induced contribution to zeroth order in the COM fluctuations given by  \eqn{\label{hdcp0}\hat{H}_\mr{DCP}^{(0)}\equiv - \sbkt{ {\mb{E}}_0 \bkt{{\mb{r}}, t} \cdot \hat{\mb{P}}_f \bkt{{\mb{r}}} +\hat{\mb{E}}_f \bkt{{\mb{r}}} \cdot {\mb{P}}_0 \bkt{{\mb{r}}, t}}.}
One can derive a corresponding drive-induced Casimir-Polder potential  in second-order perturbation theory as \cite{Fuchs18} (see Appendix\,\ref{Appdcp} for details)
\eqn{\label{udcp}
U_\mr{DCP}\bkt{\mb{r}} =&- \frac{\mu_0 \omega_0 ^2\bkt{\alpha\bkt{\omega_0 }}^2}{2}\bkt{2 n_\mr{th}\bkt{\omega_0}+1}\non\\
& \sbkt{ {\mbc{E}}_0 \bkt{\mb{r} } \cdot \re \dbar {G}_{sc} \bkt{\mb{r} , \mb{r} , \omega_0 }\cdot {\mbc{E}}_0 ^\ast\bkt{\mb{r} } }.
}
The above shift is analogous to the resonant Casimir-Polder shift for the excited state of a two-level atom \cite{SYB2, KSDEC, Fuchs18}. This can be understood as coming from a process wherein a classically-induced dipole scatters a photon off of the surface and reabsorbs it.

\subsection{Total Potential}
\label{utot}
The total potential for the classical coordinate of the particle can be written as the sum of Eq.\,\eqref{utr}, \eqref{ucp} and \eqref{udcp} as $U _\mr{Tot}\bkt{\mb{r}} = U_\mr{Tr}\bkt{\mb{r}} + U_\mr{CP}\bkt{\mb{r}}+ U_\mr{DCP}\bkt{\mb{r}}
$, which yields the classical equilibrium position $\mb{r}_0$ of the particle such that $\partial_zU_\mr{Tot}\bkt{\mb{r}}\vert_{\mb{r}_0} = 0$.

Expanding the CP and DCP Hamiltonians to second order in $\hat z$ around $\mb{r}_0$ would lead to additional corrections to the trap frequency. Assuming that the potential associated with the CP and DCP contributions are $\hat V_\mr{(D)CP} \equiv \frac{1}{2}M\Omega_\mr{(D)CP }^2 \hat{z}^2$, the total free Hamiltonian for the particle is the sum of its kinetic energy term and the harmonic trap potential given by
\eqn{\hat H_M = \frac{\hat p_z^2}{2M} + \frac{1}{2}M\Omega^2 \hat {z}^2,
}
where the total trap frequency is defined as $\Omega \equiv \sqrt{\Omega^2_\mr{Tr} + \Omega^2_\mr{CP}+\Omega^2_\mr{DCP}}$.

\section{QBM for the particle in the presence of a surface}
\label{results}

We now study  the open system dynamics of the  quantized COM motion of the particle as the system of interest, interacting with the fluctuations of the EM field as the bath. It can be seen that in the presence of an external trapping field, to the lowest order in field fluctutations, the coupling between the quantized COM motion and the EM field fluctuations arises due to the drive-induced Casimir-Polder Hamiltonian $\hat H_\mr{DCP}$.  Expanding  $\hat H_\mr{DCP}$ to first order in the COM motion fluctuations, we obtain an interaction  Hamiltonian between the quantized COM motion and the fluctuations of the field as \eqn{\label{hmf}\hat H_{MF }\bkt{t} \approx \hat {z} \hat{\mc{B}}\bkt{t},}
where $\hat{\mc{B}}\bkt{t}$ is the bath operator defined as
\begin{widetext}
\eqn{
\hat {\mc{B}}\bkt{t} =
-\sbkt{ {\mb{P}}_0 \bkt{\mb{r}_0, t} \cdot \pard{}{z}\hat{\mb{E}}_f \bkt{\mb{r}_0}+ \pard{}{z}\hat{\mb{P}}_f \bkt{\mb{r}_0} \cdot{\mb{E}}_0 \bkt{\mb{r}_0, t}+\pard{}{z}\mb{P}_0 \bkt{\mb{r}_0, t} \cdot \hat{\mb{E}}_f \bkt{\mb{r}_0}+ \hat{\mb{P}}_f \bkt{\mb{r}_0} \cdot\pard{}{z}{\mb{E}}_0 \bkt{\mb{r}_0, t}}.
}
\end{widetext}
We remark that while the last two terms in the above  with $\partial_z \mb{E}_0\bkt{\mb{r}}$ and $\partial_z \mb{P}_0\bkt{\mb{r}}$ vanish at the field intensity maxima, the equilibrium position $\mb{r}_0$ is shifted from that point due to the presence of the CP and DCP potentials.

Moving to a rotating frame of reference with respect to the total free Hamiltonian, we can write the interaction Hamiltonian (Eq.\,\eqref{hmf}) in the interaction picture as $\tilde{{H}}_{MF}(t)\equiv e^{-i\bkt{\hat H_M +\hat H_F}t}{\hat {H}}_{MF}\bkt{t}e^{i\bkt{\hat H_M +\hat H_F}t} $. In the interaction picture we can thus describe the dynamics of the COM in terms of a Born-Markov master equation as \cite{BPBook}\\

\eqn{
&\der{\hat \rho_M}{t} =\non\\
&-\frac{1}{\hbar^2} \tr_F \int_0 ^\infty \dd\tau \sbkt{\tilde {H}_{MF} (t) , \sbkt{\tilde {H}_{MF} (t-\tau), \hat{\rho}_M\bkt{t} \otimes \hat{\rho}_F}},
}

where $\hat\rho_M $ refers to the density matrix for the  quantized COM motion of the particle. Performing a trace over the EM field in the above one obtains the following QBM master equation
\begin{widetext}
\eqn{\label{qbmme}
\der{\hat \rho_M}{t} = \frac{1}{2\hbar^2} \int_0 ^\infty\dd \tau &\sbkt{i \mc{D} (\tau ) \cos\bkt{ \Omega \tau} \sbkt{\hat z, \cbkt{\hat z, \hat\rho _M}} -i \mc{D}\bkt{\tau } \frac{\sin\bkt{\Omega \tau}}{ M\Omega} \sbkt{\hat z, \cbkt{\hat {p}_z , \hat\rho_M}}\right. \non\\
&\left.- \mc{N} (\tau ) \cos\bkt{ \Omega \tau} \sbkt{\hat z, \sbkt{\hat z, \hat\rho_M }}+ \mc{N}\bkt{\tau } \frac{\sin\bkt{\Omega \tau}}{ M\Omega} \sbkt{\hat z, \sbkt{\hat {p}_z ,\hat  \rho_M}}},}
\end{widetext}
where first term  corresponds to trap frequency renormalization, the second term corresponds to dissipation or friction, the third term represents decoherence in the position basis, and the last term corresponds to momentum diffusion. 
The dissipation and noise kernels in the above master equation are given as (see Appendix\,\ref{Appdissnoise} for details of the derivation)\eqn{\label{diss}\mc{D}\bkt{\tau} &\equiv i \avg{\sbkt{\tilde{\mc{B}}\bkt{t}, \tilde{\mc{B}}\bkt{t-\tau}}}\non\\
&= 2  \hbar \int_0 ^\infty \dd \omega \, J\bkt{ \omega,  \mb{r}_0 } \sin\bkt{ \omega \tau } \cos\bkt{ \omega_0 \tau }
}

\eqn{\label{noise}&\mc{N}\bkt{\tau }\equiv\avg{\cbkt{\tilde{\mc{B}}\bkt{t}, \tilde{\mc{B}}\bkt{t-\tau}}}\non\\
&= 2 \hbar \int_0 ^\infty \dd\omega\, J\bkt{ \omega,  \mb{r}_0} \cos\bkt{\omega \tau }\cos\bkt{ \omega_0 \tau }\coth\bkt{\frac{\hbar \omega}{2k_B T}},}
where we have defined the bath operator in the interaction picture as $\tilde {\mc{B}}(t)\equiv e^{-i\bkt{\hat H_M +\hat H_F}t}\hat{\mc{B}}\bkt{t}e^{i\bkt{\hat H_M +\hat H_F}t} $. The kernels $\mc{D}\bkt{\tau }$ and $\mc{N}\bkt{\tau }$ correspond to the standard QBM dissipation and noise kernels, respectively. Here $J\bkt{\omega, \mb{r}_0}\equiv J_\mr{free}\bkt{\omega, \mb{r}_0}+J_\mr{sc}\bkt{\omega, \mb{r}_0}$ is the effective spectral density of the EM field in the presence of the surface, with the free space and scattering contributions given by

\eqn{\label{Jw}
J_\mr{free,sc}\bkt{ \omega , \mb{r}_0} = \frac{\omega^2 }{2\pi\epsilon_0 c^2 }\sbkt{\alpha\bkt{\omega_0 } + \alpha\bkt\omega}^2 g_\mr{free, sc} \bkt{\omega , \mb{r}_0},}
where we have defined
\begin{widetext}
\eqn{g_\mr{free, sc} \bkt{\omega , \mb{r}_0}\equiv&\mbc{E}_0\bkt{\mb{r}_0}\cdot\cbkt{ \partial _z\im\dbar{G}_\mr{free,sc}\bkt{\mb{r}_0, \mb{r}_0, \omega }\partial_z}\cdot\mbc{E}_0^\ast\bkt{\mb{r}_0}+\partial_z \mbc{E}_0\bkt{\mb{r}_0}\cdot  \im\dbar{G}_\mr{free,sc}\bkt{\mb{r}_0, \mb{r}_0, \omega } \cdot\partial_z\mbc{E}_0^\ast\bkt{\mb{r}_0}\non\\
 & + \mbc{E}_0\bkt{\mb{r}_0}\cdot\cbkt{ \partial _z\im\dbar{G}_\mr{free,sc}\bkt{\mb{r}_0, \mb{r}_0, \omega }}\cdot\partial_z\mbc{E}_0^\ast\bkt{\mb{r}_0}+ \partial _z\mbc{E}_0\bkt{\mb{r}_0}\cdot\cbkt{ \im\dbar{G}_\mr{free,sc}\bkt{\mb{r}_0, \mb{r}_0, \omega }\partial_z}\cdot\mbc{E}_0^\ast\bkt{\mb{r}_0}.
}
\end{widetext}

To physically interpret the  spectral density obtained above, we note the following features from Eq.\,\eqref{Jw}
\begin{itemize}
\item{The spectral density scales as the square of the induced dipole of the dielectric particle.  The part of  total spectral density that depends on $\bkt{\alpha\bkt{\omega_0}}^2$ arises due to the classically-induced dipole, corresponding to the emission and reabsorption of a photon by the classical dipole. The part of the spectral density that depends on $\bkt{\alpha\bkt{\omega}}^2$ arises from the interactions of the fluctuating dipole with its image via the classical trap field. Terms that go as $\sim { \alpha(\omega_0 )\alpha(\omega )}$ can be understood as coming from processes where a classical dipole scatters a  photon, inducing a fluctuating dipole in the medium, which in turn interacts with the classical dipole via the trap field. This can be seen from the derivation of the dissipation and noise kernels in Appendix\,\ref{Appdissnoise}.}
\item {Given that the imaginary part of the surface scattering Green's tensor $\im \dbar {G}_\mr{sc} $ corresponds to the surface loss,  the density of modes increases near a lossy surface. This indicates that a lossy surface with a large number of fluctuating degrees of freedom leads to a larger dissipation and decoherence for the quantized COM dynamics, as a result of the fluctuation-dissipation theorem.  Previously it has also been shown that surface loss leads to additional dissipation and decoherence for the internal degrees of the particle \cite{spinflip1,spinflip2, superchip, Hinds1999, Schumm2005, Lin2004,  Jones2003}.}
\item{In addition to the surface-induced modifications, there is also dissipation and decoherence due to the interaction of the particle with  the  free space EM field modes, as given by the free space Green's tensor contribution. This can be understood as arising from scattering of the classical drive photons by the particle  into free space modes.}
\end{itemize}

We can now define the dissipation and decoherence coefficients  as
\eqn{\label{Gamma}\Gamma\equiv& \frac{1}{2\hbar M \Omega} \int_0^\infty\dd \tau\, \mc{D}\bkt{\tau }\sin\bkt{\Omega\tau } \non\\
=& \frac{ \pi}{4M\Omega} \sbkt{ J\bkt{ \omega_0 + \Omega, \mb{r}_0} - J\bkt{ \omega_0 - \Omega, \mb{r}_0}}\\
\label{Lambda}
\Lambda\equiv&\frac{1}{2\hbar^2}\int_0^\infty\dd \tau\, \mc{N}\bkt{\tau }\cos\bkt{\Omega\tau } \non\\
=& \frac{ \pi}{4\hbar} \sbkt{ J\bkt{ \omega_0 + \Omega, \mb{r}_0} \coth\bkt{\frac{\hbar \bkt{\omega_0 + \Omega}}{k_B T}}\right.\non\\
&\left.+J\bkt{ \omega_0 - \Omega, \mb{r}_0}\coth\bkt{\frac{\hbar \bkt{\omega_0 - \Omega}}{k_B T}}}.
}
This shows that the dissipation and decoherence depend only on  the effective spectral density evaluated at the mechanical sideband frequencies of the driving field. It can be seen that the above expressions are analogous to those for optomechanical damping and radiation pressure induced noise \cite{OMRMP}.

This allows us to simplify the  master equation Eq.\,\eqref{qbmme} as follows
\eqn{\label{meapprox}
\der{\hat\rho_M}{t} \approx -\frac{i}{\hbar} \sbkt{H_M', \hat\rho_M} -\frac{i\Gamma}{\hbar} \sbkt{\hat z, \cbkt{\hat {p}_z , \hat\rho_M}}-\Lambda  \sbkt{\hat z, \sbkt{\hat z, \hat\rho_M }},
}
where we have defined the renormalized free Hamiltonian for the center of mass as $H_M'$ which includes the frequency renormalization due to the first term in Eq.\,\eqref{qbmme}, and ignored the momentum diffusion term \cite{BPBook}.

\section{Decoherence and quantum  friction for a dielectric nanosphere}
\label{dissdec}
As a concrete example, we now evaluate the decoherence and dissipation for a dielectric nanosphere near a planar half-space. The parameter values corresponding to the particle and surface, and simplifying assumptions are given as follows.
\subsection{Parameter values and Assumptions }
\begin{itemize}
    \item {We consider a dielectric nanosphere made of silica  with a radius $R = 72 $\,nm, and  density $\rho \approx 2000 $\,kg/$\mr{m}^3$ \cite{Magrini18}.
}
\item{The  polarizability of a dielectric nanosphere is given as \cite{Jackson}
\eqn{\dbar{\alpha}\bkt{\omega} = 3\epsilon_0 \mc{V}\sbkt{\frac{\epsilon_P \bkt{\omega}-1}{\epsilon_P \bkt{\omega}+2}} \mathbb{1},
} where $\mc{V} = \frac{4}{3}\pi R^3$ is the volume of the nanosphere, and $\epsilon_P\bkt{\omega }$ is the dielectric permittivity of the dielectric particle  described by the Drude-Lorentz model
\eqn{
\epsilon_P\bkt{\omega} = 1 + \frac{\omega_{p1}^2}{\omega_{T1}^2 - \omega^2 - i \gamma_1 \omega}+ \frac{\omega_{p2}^2}{\omega_{T2}^2 - \omega^2 - i \gamma_2 \omega}.
}
We use the parameters corresponding to fused silica as    $\omega_{p1}=1.75\times10^{14}$ Hz, $\gamma_{1}=4.28\times10^{13}$ Hz,  $\omega_{T1}=1.32\times10^{14}$ Hz, $\omega_{p2}=2.96\times10^{16}$ Hz, $\gamma_{2}=8.09\times10^{15}$ Hz, $\omega_{T2}=2.72\times10^{16}$ Hz  \cite{Hemmerich2016}.    In the present calculations we will ignore the damping and consider only the real part of the total polarizability.
}
\item{We assume the  trap field to be polarized along the $x$-axis, with a wavelength of $\lambda_0 \approx 1064 $\,$\mu$m and intensity $ I = \frac{1}{2}\epsilon_0 \mc{E}_0^2 c\approx 10^{-11} $\,Wm$^{-2}$, as used in \cite{Magrini18}.}
\item{ We assume that the particle is trapped in a harmonic potential along the $z$ axis, with a trap frequency $\Omega \approx 3$\,MHz.}
\item{Considering that the classical drive frequency is much larger than that for the mechanical trap  $\omega_0 \gg \Omega$, and $\hbar \omega_0 \gg k_B T$ one can simplify Eq.\,\eqref{Gamma} and \eqref{Lambda} to obtain the following simple expressions for the dissipation and noise 
\eqn{\label{gamma}
\Gamma &\approx \frac{\pi }{2 M } J' \bkt{\omega_0, \mb{r}_0 }\\
\label{lambda}
\Lambda &\approx \frac{\pi}{2 \hbar } J\bkt{\omega_0, \mb{r}_0 }.
}
}
\item{
For the purpose of estimation we consider in the following that the equilibrium position is roughly given by the classical trap field intensity maximum, such that we can approximate the spectral density in Eq.\,\eqref{Jw} as
\eqn{\label{Jwapprox}
 &J_\mr{free,sc}\bkt{ \omega , \mb{r}_0} \non\\
 &\approx\frac{\omega^2 }{2\pi\epsilon_0 c^2 }\sbkt{\alpha\bkt{\omega_0 } + \alpha\bkt\omega}^2\abs{\mbc{E}_0\bkt{\mb{r}_0}}^2\mc{G}_\mr{free,sc }\bkt{\omega, \mb{r}_0},
}
where we have assumed that $\partial_z \mbc{E}_0 \bkt{\mb{r}_0 }\approx 0$ and defined the free-space and scattering recoil Green's tensor as \eqn{\label{recG}\mc{G}_\mr{sc, free}\bkt{\omega, \mb{r}_0}\equiv \mb{e}_x\cdot\partial _z \im{ \dbar{G}_\mr{free,sc}\bkt{\mb{r}_0, \mb{r}_0, \omega }}\partial_z \cdot \mb{e}_x,}
where we have assumed the trap field to be polarized along the $x$-axis.
}
\end{itemize}

\subsection{Surface properties}
We calculate the influence of the free space and surface scattered EM field given by the the imaginary part of the Green's tensor as follows.
\begin{figure}
        \includegraphics[width = 3.6 in]{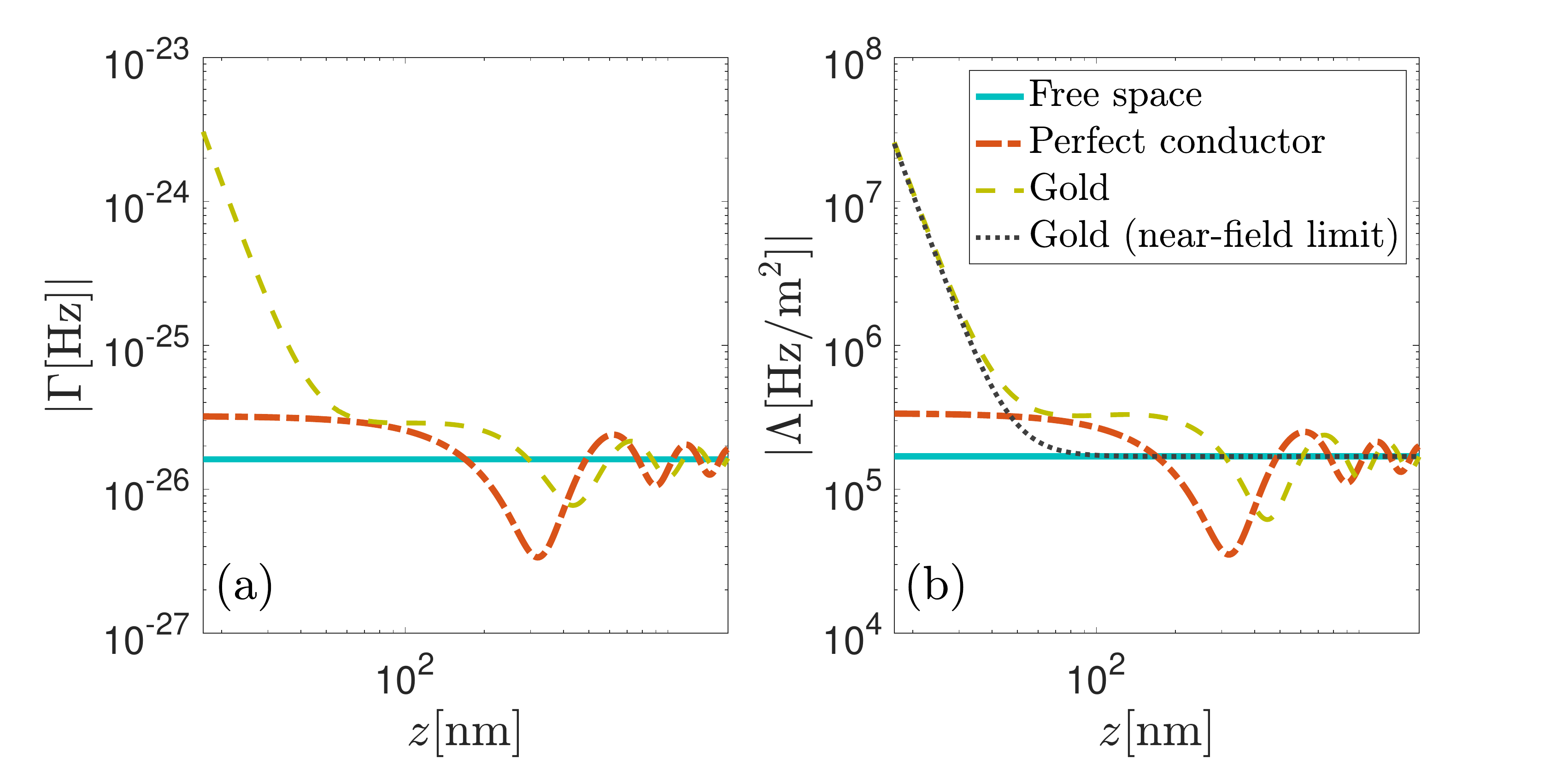}
    \caption{(a) Dissipation and (b) decoherence of a silica nanosphere near  perfect conductor and gold surfaces. The free space dissipation and decoherence is depicted by the blue solid line in both the plots.  The dotted line in (b) denotes the near-field asymptotic expression  for decoherence of the particle near a metal surface as given by Eq.\,\eqref{Lmet}. The gold surface is described by the Drude model with a plasma frequency $\omega_p\approx1.37\times10^{16}$\,Hz (9 eV), and loss parameter $\gamma \approx 5.31\times10^{13}$\,Hz (35 meV) \cite{Pirozhenko2006}.} 
    \label{Fig2}
\end{figure}
\subsubsection{Free space}
Observing that the free space recoil Green's tensor is $\mc{G}_{\mr{free}}\bkt{\omega,\mb{r}_0} = \frac{\omega^3}{15\pi c^3 }$, using Eq.\,\eqref{Jwapprox} the contribution to the spectral density due to the free space EM field modes is given by 
   
  \eqn{\label{Jwfree}
  J_\mr{free}\bkt{\omega} = \frac{2\hbar \omega^2 \gamma_0\bkt{\omega}}{5\pi c^2 },
  }
  where we have defined  \eqn{\gamma_0\bkt{\omega} \equiv \frac{ \sbkt{\alpha(\omega_0 ) + \alpha(\omega )}^2 \abs{\mbc{E}_0}^2 \omega^3 }{12\pi \epsilon_0 \hbar c^3},} analogous to the dissipation rate of a dipole of strength $ {d}\equiv \sbkt{\frac{\alpha\bkt{\omega_0 } + \alpha\bkt{\omega }}{2}} \abs{\mbc{E}_0}$ interacting with the vacuum EM field. It can be seen from Eq.\,\eqref{lambda} that this yields a decoherence rate of
  \eqn{\label{Lfree} \Lambda_\mr{free} = \frac{\omega_0^2\gamma_0 \bkt{\omega_0} }{5 c^2 },}
which corresponds to the position decoherence of the particle arising from the scattering of drive photons into free space modes.
\subsubsection{Perfect conductor}
For a perfectly conducting planar surface (with the Fresnel coefficients $r_p = 1$ and $r_s  = -1$) the scattering part of the recoil Green's tensor is given as (see Eq.\,\eqref{Gzzpc}),
\eqn{
&\mc{G}_{\mr{sc}, \mr{pc}}\bkt{\omega,\mb{r}_0}=\non\\
&\frac{\omega^3}{32{\pi \tilde {z}^5}c^3}\sbkt{\left(\tilde {z}^2-1\right) \cbkt{ 6 \tilde {z}\cos (2 \tilde {z})+\left(4 \tilde {z}^2-3\right)\sin (2 \tilde {z})}},
}
where $\tilde z \equiv k_0 z $ is the dimensionless distance of the particle from the surface. In the near-field limit $\tilde z \ll 1$, we find that $\mc{G}_\mr{sc,  \mr{pc}}^{ \mr{NR}}\bkt{\omega, \mb{r}_0} \approx \frac{\omega^3}{15\pi c^3}$, where NR stands for non-retarded regime. This yields the density of modes in the subwavelength limit as
\eqn{\label{Jwpc}
J_\mr{pc}^{\mr{NR}}\bkt{\omega, \mb{r}_0} \approx J_\mr{free} \bkt{\omega, \mb{r}_0}+ J_\mr{sc}^\mr{NR}\bkt{\omega, \mb{r}_0}\approx\frac{4\hbar \omega^2 \gamma_0\bkt{\omega}}{5\pi c^2 } .
}
Comparing with Eq.\,\eqref{Jwfree}, we see that the density of modes is twice that of the free space, which can be  understood as a sum of the field radiated by the dipole and its image. 

The corresponding localization parameter is given as

\eqn{
\Lambda_\mr{pc}^\mr{NR}  \approx \frac{2\omega_0 ^2\gamma_0 \bkt{\omega_0 }}{5c^2}.
}
We note that in the absence of surface losses the decoherence of the particle in the near-field regime is independent of its distance from the surface, as can be seen from Fig.\,\ref{Fig2}\,(b).

\subsubsection{Metal}
For a metal surface we assume the permittivity function to be given by the Drude model
\eqn{\label{Drude}
\epsilon_S \bkt{\omega} = 1 - \frac{\omega_p ^2}{ \omega ^2 + i \omega \gamma}.
}

The scattering recoil Green's tensor is given as (see Eq.\,\eqref{Gzzpc})
\eqn{
    \mc{G}_{\mr{sc}, \mr{met}}\bkt{\omega, \mb{r}_0}=\frac{1}{8\pi }\im&\sbkt{\int_0 ^\infty \dd k_\parallel\,k_\parallel\kappa_\perp e^{-2 \kappa_\perp z }\right.  \non\\
     &\left. \cbkt{ r_p \bkt{ \kappa_\perp ,k_0 } \frac{\kappa_\perp ^2}{k_0 ^2}+ r_s\bkt{ \kappa_\perp , k_0 } }},
}
where the Fresnel coefficients are as given by Eq.\,\eqref{fresnel}.

  In the near-field limit the scattering recoil Green's tensor can be simplified to $\mc{G}^{ \mr{NR}}_{\mr{sc}, \mr{met}}\bkt{\omega, \mb{r}_0} \approx\frac{3 \omega^3}{ 32 \pi \tilde {z}^5c^3} \im\sbkt{ \frac{\epsilon_S\bkt{\omega} - 1}{\epsilon_S\bkt{\omega} + 1}}$, yielding a density of modes near a metal surface as 
\eqn{
J_{\mr{met}}^{\mr{NR}}\bkt{\omega, \mb{r}_0}\approx \frac{9\hbar\omega^2 }{16\pi c^2 \tilde z ^5} \im\sbkt{ \frac{\epsilon_S\bkt{\omega} - 1}{\epsilon_S\bkt{\omega} + 1}}\gamma_0 \bkt{\omega}.
}
One can thus write the decoherence of the particle near a metal half-space as 
\eqn{\label{Lmet}
\Lambda_\mr{met}^\mr{NR} \approx \frac{9 k_0^2}{32\tilde z ^5 } \im \sbkt{ \frac{\epsilon_S \bkt{\omega_0 } - 1}{\epsilon_S\bkt{\omega_0 } + 1}}\gamma_0 \bkt{\omega_0 }\approx \frac{3}{4 z^2}\gamma_\mr{sc}\bkt{\mb{r}_0},
} where  $\gamma_\mr{sc}\bkt{\mb{r}_0} \approx \frac{3}{8\tilde z^3}\im \sbkt{\frac{\epsilon_S \bkt{\omega_0 }-1}{\epsilon_S \bkt{\omega_0 }+1}}\gamma_0 \bkt{\omega_0 }$ is surface-modified photon scattering rate (see Eq.\,\eqref{gammasc}). As seen from Fig.\,\ref{Fig2}\,(b), the decoherence of the particle's COM in the near field regime is  well-approximated by the above expression.

\subsection{Correspondence to collisional model of decoherence}
We note that the deocherence term in the master equation Eq.\,\eqref{meapprox} is of the position localization decoherence (PLD) form \cite{MaxBook}. To understand this, we note that a similar form of  decoherence term can also be obtained from a collisional model of decoherence, wherein the system in consideration is bombarded by individual scatterers from the environment. As each scattering bath particle interacts  with the system via a local interaction and gets correlated, it acquires some information about the system's position as a result. Thus, upon tracing out the bath, the system exhibits  decoherence in the position basis. Particularly in the  limit where the  scatterer has a much longer de Broglie wavelength compared to the coherence length scale of the system one obtains a decoherence term as in Eq.\,\eqref{meapprox} \cite{JoosZeh, MaxBook}. This correspondence  in the decoherence dynamics  from two different models suggests that the decoherence of a particle near a surface arises due to scattering of virtual photons off of the surface. 

We note that the decoherence rate due to scattering of photons in free space goes as $\Lambda_\mr{free}\sim k_\mr{eff}^2 \gamma_\mr{eff}$ (see Eq.\,\eqref{Lfree}), where $k_\mr{eff}$ refers to an effective wavevector for the scattered photon and $\gamma_\mr{eff}$ is the rate of scattering. Considering that $\Lambda_\mr{met}^\mr{NR}\sim \gamma_\mr{sc}\bkt{\mb{r}_0 }/z^2$, we deduce from Eq.\,\eqref{Lmet} that the virtual photons inducing decoherence have an effective de Broglie wavelength $\sim k_\mr{eff}^{-1}\sim z$ that scales as the distance of the particle from the surface. We remark that a similar effective de Broglie wavelength was previously also derived in \cite{KSDEC} in the context of recoil heating of a driven atom near a surface.

\section{Discussion}
\label{discussion}
To summarize, we have derived a quantum Brownian motion master equation for the quantized center of mass motion of a dielectric particle trapped near a surface.  Considering the particle to be trapped with an external classical field, we find that there are three different contributions to the total potential seen by the particle -- a classical trap potential,  Casimir-Polder potential and driven CP potential, as illustrated in Fig.\,\ref{schematic}(b). Taken together, these potentials lead the particle to be trapped close to the first intensity maxima of the standing wave potential formed by the classical field (see Section\,\ref{utot}). The interaction between the quantized COM motion and the fluctuations of the EM field to the lowest order is described by a linear expansion of the driven CP potential (Eq.\,\eqref{hdcp}) about the equilibrium position. Tracing out the EM field fluctutations, we arrive at a second-order Born-Markov master equation describing the dissipative dynamics of the particle's COM as given by  Eq.\,\eqref{qbmme}. The resulting dynamics is governed by a  quantum Brownian motion master equation, with an effective spectral density that is determined by the polarizability of the particle, properties of the surface, and the strength of the external trapping field (Eq.\,\eqref{Jw}). The dissipation and decoherence that arise as a result can be understood as coming from the classically induced dipole scattering field fluctuations, and the fluctuating dipole scattering the drive photons. We then estimate the decoherence and dissipation for a dielectric nanosphere near different surfaces, and find that the quantized COM decoherence and dissipation increase in the presence of a lossy medium (Fig.\,\ref{Fig2}). We further illustrate a correspondence between the resulting decoherence and that from the collisional model in the long-wavelength limit \cite{JoosZeh, MaxBook}. 

Comparing the resulting decoherence due to surface fluctuations  with that arising from other sources as a benchmark, we observe that surface-induced decoherence can potentially pose a fundamental limit for preparing a dielectric particle in macroscopic COM quantum states. It can be seen from  Appendix\,\ref{Appdecother} that the decoherence due to background gas scattering and blackbody radiation can be reduced significantly  by going to lower pressures and temperatures, respectively. In the present analysis we have derived the spectral density that governs the surface-modifications to fluctuation phenomena for the quantized COM of a particle. This could allow one to systematically modify the surface properties and drive strength  in order to mitigate the surface-induced dissipation and decoherence.  As quantum optical systems are being increasingly miniaturized, and mesoscopic quantum components being regularly interfaced with surfaces and waveguides at nanoscales,  our results provide  new insights into tailoring fluctuation phenomena in these regimes \cite{Arod11, Woods16, KS2017, KS2018}.
\acknowledgements{We are grateful to Peter W. Milonni, Bei-Lok Hu, Pierre Meystre, Oriol Romero-Isart, Wade D. Hodson,  Uro\v{s} Deli\'{c}, Lorenzo Magrini, Sungkun Hong, and Markus Aspelmeyer for insightful discussions. K.S. would like to thank the Aspelmeyer group for graciously hosting her visit at the University of Vienna where part of this work was carried out. Y.S. acknowledges support from the Los Alamos National Laboratory ASC Beyond Moore's Law project and the LDRD program. The authors enjoyed the hospitality of IHoP (International House of Physicists) during the foundational stages of this work.}
\appendix

\section{Medium-assisted EM field}
\label{appendixa}
Using the macroscopic QED formalism \cite{SYB1, SYB2}, the Hamiltonian for the vacuum EM field in the presence of the surface can be written as
\eqn{\label{hv}
H_F =\sum_{ \lambda = e,m}\int \dd^3 r \int \dd\omega\,\hbar\omega\, \hat{\mb{f}}^\dagger_\lambda\bkt{\mb{r}, \omega}\cdot\hat{\mb{f}}_\lambda\bkt{\mb{r}, \omega},
}
with $\hat{\mb{f}}^\dagger_\lambda\bkt{\mb{r}, \omega}$ and ${\hat{\mb{f}}_\lambda\bkt{\mb{r}, \omega}}$ as the bosonic creation and annihilation operators respectively that take into account the presence of the media. Physically these can be understood as the ladder operators corresponding to the noise polarization ($\lambda = e$) and magnetization  ($\lambda = m$) in the medium-assisted EM field, at frequency $\omega$, created or annihilated at position $\mb{r}$. The medium-assisted bosonic operators obey the canonical commutation relations \eqn{\sbkt{ \hat{\mb{f}}_{\lambda}\bkt{\mb{r}, \omega}, \hat{\mb{f}}_{\lambda'}\bkt{\mb{r}', \omega'} } = \sbkt{ \hat{\mb{f}}^{\dagger}_{\lambda}\bkt{\mb{r}, \omega}, \hat{\mb{f}}^\dagger_{\lambda'}\bkt{\mb{r}', \omega'} }=0,\\
\sbkt{ \hat{\mb{f}}_{\lambda}\bkt{\mb{r}, \omega}, \hat{\mb{f}}^\dagger_{\lambda'}\bkt{\mb{r}', \omega'} } = \delta_{\lambda\lambda'}\delta\bkt{\mb{r} - \mb{r}'}\delta\bkt{\omega - \omega'}.}

The electric and magnetic field operators evaluated  at the position of the particle are given as 
\begin{widetext}
\eqn{\label{Era} \hat{\mb{E}}_f\bkt{\mb{r}_0}=& \sum_{ \lambda = e,m}\int \dd^3 r\int\dd\omega
 \sbkt{\dbar {G}_\lambda \bkt{\mb{r}_0, \mb{r}, \omega}\cdot \hat{\mb{f}}_{\lambda}\bkt{\mb{r},  \omega} + \text{H.C.}},
\text{ and}\\
\label{Bra}
\hat{\mb{B}}_f\bkt{\mb{r}_0}=& \sum_{ \lambda = e,m}\int \dd^3 r\int\dd\omega \, \left[\bkt{-\frac{i}{\omega}}\sbkt{\mb{\nabla}\times\dbar {G}_\lambda \bkt{\mb{r}_0, \mb{r}, \omega}}\cdot
 \hat{\mb{f}}_{\lambda}\bkt{\mb{r},  \omega} + \text{H.c.}\right]}
 \end{widetext}
 respectively, where \eqn{[{\bf \nabla}\times\bar{\bar {G}}_\lambda({\bf r},{\bf r}'\omega)]_{il} = \epsilon_{ijk}\partial_{r_j} [\bar{\bar{G}}_\lambda(\mb{r},\mb{r}',\omega)]_{kl}.}
 
 The coefficients $\dbar{G}_\lambda\bkt{\mb{r}_1,\mb{r}_2,\omega}$  are defined as 
\eqn{\label{Ge}\dbar{G}_e \bkt{\mb{r},\mb{r}',\omega}=& i\frac{\omega^2}{c^2} \sqrt{\frac{\hbar}{\pi\epsilon_0}\im[\epsilon \bkt{\mb{r}',\omega}]} \dbar{G}\bkt{\mb{r},\mb{r}',\omega},}
\eqn{\label{Gm}\dbar{G}_m \bkt{\mb{r},\mb{r}',\omega}=& \frac{i\omega}{c} \sqrt{\frac{\hbar}{\pi\epsilon_0}\frac{\im [\mu \bkt{\mb{r}', \omega}]}{\abs{\mu\bkt{\mb{r}',\omega}}^2}}\sbkt{\nabla'\times \dbar{G}\bkt{\mb{r}',\mb{r},\omega}}^T,}
with $\epsilon(\mb{r},\omega)$ and $\mu(\mb{r},\omega)$ as the space-dependent  permittivity and permeability, and $\dbar{G}\bkt{\mb{r},\mb{r}',\omega}$ as the Green's tensor for a point dipole near a  surface \cite{SYB1,SYB2}.  The Green's tensor is defined as the solution to the Helmholtz equation in the presence of the boundary conditions
\eqn{\label{Gdef}
\mb{\nabla}\times \mb{\nabla}\times \dbar{G} \bkt{\mb{r},\mb{r}', \omega } - \frac{\omega^2}{c^2}\epsilon\bkt{\mb{r}, \omega}\mu\bkt{\mb{r}, \omega} \dbar{G}\bkt{\mb{r},\mb{r}', \omega}& \non\\
= \delta \bkt{\mb{r}-\mb{r}'} \mathbb{I}&.
}

The total Green's tensor can be expressed as  
\eqn{ \label{Gtot}\dbar G\bkt{\mb{r}_1,\mb{r}_2,\omega} = \dbar G_\mr{free}\bkt{\mb{r}_1,\mb{r}_2,\omega}+\dbar G_\mr{sc}\bkt{\mb{r}_1,\mb{r}_2,\omega} ,}
where $\dbar{ G}_\mr{free}\bkt{\mb{r}_1,\mb{r}_2,\omega}$  and $\dbar{G}_\mr{sc}\bkt{\mb{r}_1,\mb{r}_2,\omega}$ refer to the free space and scattering components of the total Green's tensor.
\section{Scattering Green's tensor near a planar surface}
\label{AppGreen}
For a point dipole located at the position $\mb{r}_1$ near an infinite planar half-space, one can  write the scattering  Green's tensor as \cite{SYB1}
\begin{widetext}
\eqn{\label{greensc}
  \dbar{G}_{\mr{sc}}\bkt{\mb{r}_1,\mb{r}_2,\omega} &= \frac{1}{8\pi} \int_0^\infty \dd k_\parallel \frac{k_\parallel}{ \kappa_\perp } e^{- \kappa_\perp \bkt{z_1+z_2}}\left[\left(\begin{array}{ccc}
    J_0(k_\parallel x_{12}) +J_2(k_\parallel x_{12}) &0 &0 \\
    0 &J_0(k_\parallel x_{12}) -J_2(k_\parallel x_{12})  &0\\
    0& 0&0
\end{array}\right)r_s\right.\non\\
&\left.+\frac{c^2}{\omega^2} \bkt{\begin{array}{ccc}
    \kappa_\perp^2 \sbkt{J_0(k_\parallel x_{12}) -J_2(k_\parallel x_{12}) }&0 &2k_\parallel \kappa_\perp J_1 (k_\parallel x_{12}) \\
    0 &\kappa_\perp^2 \sbkt{J_0(k_\parallel x_{12}) +J_2(k_\parallel x_{12}) }  &0\\
    -2k_\parallel \kappa_\perp J_1 (k_\parallel x_{12})& 0&2 k_\parallel^2 J_0 (k_\parallel x_{12})
\end{array}}r_p \right],
}
\end{widetext}
with $\abs{{\mb{r}_1-\mb{r}_2} }= r$, ${({\mb{r}_1+\mb{r}_2}) \cdot \mb{e}_z}= \bkt{z_1+z_2}$, and we have defined the relative coordinate vector between the points $\mb{r}_1$ and $\mb{r}_2$ as ${\frac{{\mb{r}_1-\mb{r}_2}}{\abs{\mb{r}_1-\mb{r}_2}}\equiv \bkt{\frac{x_{12}}{r},0,\frac{z_1 - z_2}{r}}^\mr{T}.}$ Here $r_{s,p}$ are the Fresnel reflection coefficients for the $s$ and $p$ polarizations reflecting off the surface, and $\kappa_\perp^2 = -k^2 +k_\parallel^2$, where $k = \omega/c$. Assuming that the medium can be treated as homogeneous and isotropic, and can be well-described in terms of its bulk optical properties at the length scales of the particle-surface separations, we can consider that all the information about the surface material is accounted for in the following Fresnel  reflection coefficients
\eqn{\label{fresnel}
r_p\bkt{\kappa_\perp, \omega} &=  \frac{\epsilon\bkt{\omega}\kappa_\perp-\sqrt{-\bkt{\epsilon\bkt{\omega}\mu\bkt{\omega}-1}k^2+\kappa_\perp^2}}{\epsilon\bkt{\omega}\kappa_\perp+\sqrt{-\bkt{\epsilon\bkt{\omega}\mu\bkt{\omega}-1}k^2+\kappa_\perp^2}},\non\\
r_s \bkt{\kappa_\perp, \omega}&= \frac{\mu\bkt{\omega}\kappa_\perp-\sqrt{-\bkt{\epsilon\bkt{\omega}\mu\bkt{\omega}-1}k^2+\kappa_\perp^2}}{\mu\bkt{\omega}\kappa_\perp+\sqrt{-\bkt{\epsilon\bkt{\omega}\mu\bkt{\omega}-1}k^2+\kappa_\perp^2}}.
}

In the non-retarded limit $(\tilde z\ll1)$, one can expand the Fresnel coefficients in Eq.\,\eqref{fresnel} to lowest order in $\abs{\sqrt{\epsilon(\omega)-1}\omega/(\kappa_\perp c)}$ as
\eqn{\label{rpapprox}
r_p\bkt{\kappa_\perp, \omega}&\approx\frac{\epsilon\bkt{\omega}-1}{\epsilon\bkt{\omega}+1} + \frac{\epsilon\bkt{\omega}\bkt{\epsilon\bkt{\omega}-1}}{\bkt{\epsilon\bkt{\omega}+1}^2}\frac{\omega^2}{\kappa_\perp^2c^2},\\
\label{rsapprox}
r_s\bkt{\kappa_\perp, \omega}&\approx \frac{1}{4}\bkt{\epsilon\bkt{\omega}-1}\frac{\omega^2}{\kappa_\perp^2c^2}.
}

For coincident points $(\mb{r}_1 = \mb{r}_2 = \mb{r}_0)$, one can  write the scattering Green's tensor as \cite{SYB1}
\begin{widetext}
\eqn{\label{green}
 \dbar{G}_{\mr{sc}}\bkt{{\bf r}_0,{\bf r}_0,\omega} &= \frac{1}{8\pi} \int_0^\infty\frac{\dd k_{\parallel}\,k_{\parallel} }{\kappa_\perp}e^{-2\kappa_\perp z}\left[r_p\bkt{\kappa_\perp, \omega}\frac{c^2}{\omega^2} \bkt{\begin{array}{ccc}
    \kappa_\perp^2 &0 &0 \\
    0 &\kappa_\perp^2 &0\\
    0& 0& 2k_\parallel^2
\end{array}}+r_s\bkt{\kappa_\perp, \omega} \bkt{\begin{array}{ccc}
    1 &0 &0 \\
    0 &1 &0\\
    0& 0& 0
\end{array}} \right], 
}
\end{widetext}

The double $z$-derivative of the $xx$ component of the recoil Green's tensor (see Eq.\,\eqref{recG}) is then given as
  \eqn{\label{Gzzpc}
 &   \partial_z\im \dbar{G}_{\mr{sc}}^{xx}\bkt{\mb{r}_0 , \mb{r}_0 , \omega_0}\partial_z = \frac{1}{8\pi }\im\sbkt{ \int_0 ^\infty \frac{\dd k_\parallel\,k_\parallel}{ \kappa_\perp }\kappa_\perp ^2e^{-2 \kappa_\perp z } \right.\non\\
&  \quad\quad\quad\quad  \quad\quad   \left.\cbkt{ r_p \bkt{ \kappa_\perp ,k_0 } \frac{\kappa_\perp ^2}{k_0 ^2}+ r_s\bkt{ \kappa_\perp , k_0 } }}.
  }
The free space Green's tensor between the points $\mb{r}_1$ and $\mb{r}_2$ is given as 

\eqn{
\label{greenfree}
&\dbar{G}_{\mr{free}} \bkt{\mb{r}_1,\mb{r}_2, \omega} =\non\\
&-\frac{ e^{i k r}}{4\pi k^2 r^3} \bkt{\begin{array}{ccc}
     f\bkt{k r} - h\bkt{k r}\frac{x_{12}^2}{r^2}& 0&0 \\
     0&f\bkt{k r}&0 \\
0   &0 &f\bkt{k r}
\end{array}} .}

where $f\bkt{x}\equiv1-ix-x^2$, 
$h\bkt{x}\equiv3-3ix-x^2$.

\section{Derivation of the driven Casimir-Polder potential}
\label{Appdcp}

Using time-dependent second-order perturbation theory, we can define the energy correction and the modification to the dissipation rate of the system   arising due to the driven CP Hamiltonian $\hat{H}_\mr{DCP}^{(0)}\bkt{t}$ (see Eq.\,\eqref{hdcp0}) as 
$U_\mr{DCP }\bkt{\mb{r} } = \hbar \re \chi$, and 
$\gamma_\mr{sc}\bkt{\mb{r} } = -\im \chi$, where

\begin{widetext}
\eqn{
\chi = &- \frac{i}{\hbar ^2 }\avg{ \int_0 ^\infty \dd\tau  \tilde{H}_\mr{DCP}^{(0)}(t) \tilde {H}_\mr{DCP}^{(0)}(t- \tau)}_F\\
 = & - \frac{i}{\hbar^ 2} \left\langle \int_0 ^\infty  \dd\tau \cbkt{\mb{P}_0 \bkt{\mb{r}, t} \cdot \tilde{\mb{E}}_f \bkt{\mb{r}, t} + \tilde{\mb{P}}_f \bkt{\mb{r}, t} \cdot \mb{E}_0 \bkt{\mb{r}, t}}\right.\non\\
 &\cbkt{\mb{P}_0 \bkt{\mb{r}, t-\tau} \cdot \tilde{\mb{E}}_f \bkt{\mb{r}, t-\tau} + \tilde{\mb{P}}_f \bkt{\mb{r}, t-\tau} \cdot \mb{E}_0 \bkt{\mb{r}, t-\tau} }\biggr>_F,
 }
 \end{widetext}
 where we have defined the electric field and polarization fluctuation operators in the interaction picture as $\tilde{\mc{O}}(t)\equiv e^{-i\bkt{\hat H_M +\hat H_F}t}\hat{\mc {O}}e^{i\bkt{\hat H_M +\hat H_F}t} $ \cite{Sakurai}. The average is taken over the thermal state of the field. We note that the shifts and decay rates  are consistent with those derived via the second-order Born-Markov master equation \cite{BPBook}.

We further divide the above into 4 separate terms as
\eqn{
\bkt{\mr{I}}&\equiv -\frac{i}{\hbar^2}\left\langle \int_0 ^\infty \dd\tau \cbkt{\mb{P}_0 \bkt{\mb{r}, t} \cdot \tilde{\mb{E}}_f \bkt{\mb{r}, t}}\right.\non\\
&\cbkt{\mb{P}_0 \bkt{\mb{r}, t-\tau } \cdot \tilde{\mb{E}}_f \bkt{\mb{r}, t-\tau }}   \biggr>_F}
\eqn{
\bkt{\mr{II}}&\equiv -\frac{i}{\hbar^2}\left\langle \int_0 ^\infty \dd\tau \cbkt{\mb{P}_0 \bkt{\mb{r}, t} \cdot \tilde{\mb{E}}_f \bkt{\mb{r}, t}}\right.\non\\
&\cbkt{ \tilde{\mb{P}}_f \bkt{\mb{r}, t-\tau }\cdot \mb{E}_0 \bkt{\mb{r}, t-\tau }}  \biggr>_F}
\eqn{
\bkt{\mr{III}}&\equiv -\frac{i}{\hbar^2}\left\langle \int_0 ^\infty \dd\tau\cbkt{ \tilde{\mb{P}}_f \bkt{\mb{r}, t}\cdot \mb{E}_0 \bkt{\mb{r}, t}  }\right.\non\\
&\cbkt{\mb{P}_0 \bkt{\mb{r}, t-\tau } \cdot \tilde{\mb{E}}_f \bkt{\mb{r}, t-\tau }}\biggr>_F}
\eqn{
\bkt{\mr{IV}}&\equiv -\frac{i}{\hbar^2}\left\langle \int_0 ^\infty \dd\tau \cbkt{ \tilde{\mb{P}}_f \bkt{\mb{r}, t}\cdot \mb{E}_0 \bkt{\mb{r}, t}}\right.\non\\
&\cbkt{ \tilde{\mb{P}}_f \bkt{\mb{r}, t-\tau }\cdot \mb{E}_0 \bkt{\mb{r}, t-\tau } }\biggr>_F,
}

such that $\mr{\chi =(I) + (II) + (III)  + (IV)  }$. Let us consider the first term as follows
\begin{widetext}
\eqn{
&\bkt{\mr{I}}= -\frac{i}{\hbar^2}\tr_F \sbkt{ \int_0 ^\infty \dd\tau \cbkt{\mb{P}_0 \bkt{\mb{r}, t} \cdot \tilde{\mb{E}}_f \bkt{\mb{r}, t}}\cbkt{\mb{P}_0 \bkt{\mb{r}, t-\tau } \cdot \tilde{\mb{E}}_f \bkt{\mb{r}, t-\tau }}\hat \rho_F  }\\
& = -\frac{i}{4\hbar^2}\tr_F \sbkt{ \int_0 ^\infty \dd\tau\cbkt{\alpha\bkt{\omega_0 }\mbc{E}_0 \bkt{\mb{r}} e^{- i \omega_0 t}+ \alpha\bkt{\omega_0 }\mbc{E}_0 ^\ast\bkt{\mb{r}} e^{i \omega_0 t}}\cdot \right.\non\\
&\left.\cbkt{\int\dd\omega_1 \sum_{\lambda_1 = e,m} \int\dd^3 r_1 \bkt{\dbar {G}_{\lambda_1} \bkt{ \mb{r},\mb{r}_1, \omega_1}\cdot \hat{\mb{f}}_{\lambda_1}\bkt{\mb{r}_1,  \omega_1}e^{-i \omega_1 t} + \hat{\mb{f}}^\dagger_{\lambda_1}\bkt{\mb{r}_1,  \omega_1}\cdot\dbar {G}_{\lambda_1}^\dagger \bkt{ \mb{r},\mb{r}_1, \omega_1} e^{i \omega_1 t}}}\right.\non\\
&\left.\cbkt{\int\dd\omega_2 \sum_{\lambda_2 = e,m} \int\dd^3 r_2 \bkt{\dbar {G}_{\lambda_2} \bkt{ \mb{r}, \mb{r}_2,\omega_1}\cdot \hat{\mb{f}}_{\lambda_2}\bkt{\mb{r}_2,  \omega_2}e^{-i \omega_2 (t- \tau)} + \hat{\mb{f}}^\dagger_{\lambda_2}\bkt{\mb{r}_2,  \omega_2}\cdot\dbar {G}_{\lambda_2}^\dagger \bkt{ \mb{r},\mb{r}_2, \omega_2} e^{i \omega_2 (t- \tau)}}}\cdot \right.\non\\
&\left.\cbkt{\alpha\bkt{\omega_0 }\mbc{E}_0 \bkt{\mb{r}} e^{- i \omega_0 (t-\tau)}+ \alpha\bkt{\omega_0 }\mbc{E}_0^\ast \bkt{\mb{r}}e^{i \omega_0 (t- \tau)}} \hat \rho_F}\\
& = -\frac{i}{4\hbar^2}\int _0 ^\infty \dd\tau\bkt{\alpha\bkt{\omega_0 }}^2 \sbkt{\mbc{E}_0 \bkt{\mb{r}} e^{- i \omega_0 t}+ \mbc{E}_0 ^\ast\bkt{\mb{r}} e^{i \omega_0 t}}\cdot\sbkt{\int \dd\omega \sum_{\lambda = e,m}\int\dd^3 r'\cbkt{\dbar {G}_{\lambda} \bkt{\mb{r},\mb{r}',  \omega}\cdot\dbar {G}^\dagger_{\lambda} \bkt{ \mb{r}, \mb{r}',\omega} }\right.\non\\
&\left. \cbkt{\bkt{n_\mr{th }\bkt{\omega}+1}e^{-i \omega \tau } + n_\mr{th}\bkt{\omega}e^{i \omega \tau }} }\cdot  \sbkt{\mbc{E}_0 \bkt{\mb{r}} e^{- i \omega_0 (t-\tau)}+ \mbc{E}_0 ^\ast\bkt{\mb{r}} e^{i \omega_0 (t-\tau)}} }
\end{widetext}
 where in the second step we take an average over the field density matrix $\hat\rho_F =\hat \rho_\mr{th}$, such that \eqn{&\tr_F\sbkt{\hat{\mb{f}}_{\lambda_1}^\dagger\bkt{\mb{r}_1, \omega_1}\cdot \hat{\mb{f}}_{\lambda_2}\bkt{\mb{r}_2, \omega_2}\rho_F} \non\\
 &= n_{\mr{th}}\bkt{\omega_1}\delta_{\lambda_1 ,\lambda_2 }\delta\bkt{\mb{r}_1 - \mb{r}_2}\delta\bkt{{\omega}_1 - {\omega}_2}.}  Physically this is equivalent to saying that the virtual excitations  of the EM field emitted and absorbed by the particle occur at the same frequency, position and space coordinate, and have an average number expectation value of $ n_\mr{th}\bkt{\omega}$. We can further simplify $(\mr{I})$ as
 \begin{widetext}
 \eqn{
\bkt{\mr{I}} =& -\frac{i}{4\hbar^2}\int \dd\omega \frac{\mu_0\omega^2 \bkt{\alpha\bkt{\omega_0 }}^2}{\pi\hbar}\int _0 ^\infty \dd\tau \sbkt{\mbc{E}_0 \bkt{\mb{r}} \cdot\im \dbar{G}\bkt{\mb{r}, \mb{r},\omega} \cdot \mbc{E}_0 ^\ast\bkt{\mb{r}}e^{- i \omega_0 \tau}  \right.\non\\
&\left.+\mbc{E}_0 ^\ast\bkt{\mb{r}}\cdot\im \dbar{G}\bkt{\mb{r}, \mb{r},\omega} \cdot \mbc{E}_0 \bkt{\mb{r}} e^{i \omega_0 \tau} \cbkt{\bkt{n_\mr{th }\bkt{\omega}+1} e^{-i \omega \tau } + n_\mr{th}\bkt{\omega}e^{i \omega \tau }} },\\
 =& -i\frac{ \bkt{2n_\mr{th }\bkt{\omega_0}+1}{\mu_0\omega_0^2}\bkt{\alpha\bkt{\omega_0 }}^2}{4\hbar} \sbkt{\mbc{E}_0 \bkt{\mb{r}} \cdot\im \dbar{G}\bkt{\mb{r}, \mb{r},\omega_0} \cdot \mbc{E}_0 ^\ast\bkt{\mb{r}} }\non\\
&-\frac{ \bkt{2n_\mr{th }\bkt{\omega_0}+1}\bkt{\alpha\bkt{\omega_0 }}^2}{4\hbar}\frac{\mu_0\omega_0^2}{2} \sbkt{\mbc{E}_0 \bkt{\mb{r}} \cdot\re \dbar{G}\bkt{\mb{r}, \mb{r},\omega_0} \cdot \mbc{E}_0 ^\ast\bkt{\mb{r}} },
}
\end{widetext}
 where in the first step we have made the rotating wave approximation and used  the fluctuation-dissipation relation \cite{SYB1}
 \eqn{\sum_{\lambda = e,m}\int \dd^3 r'\, \dbar{G}_\lambda \bkt{\mb{r},\mb{r}', \omega}& \dbar{G}^\dagger_\lambda \bkt{\mb{r},\mb{r}', \omega}\non\\
& = \frac{\hbar\mu_0\omega^2}{\pi} \im \dbar{G}\bkt{\mb{r}, \mb{r},\omega}  .}
In the second step while performing the integral over $\tau $, we note that $\int_0 ^\infty \dd\tau e^{i \tau x } = \pi \delta \bkt{x} + i \mc{P}\bkt{\frac{1}{x}}$. To evaluate the principal value term, we make a contour integral over the first and the second quadrants of the upper half complex plane.

We can similarly simplify the remaining terms to find that $\mr{(I) = (II) = (III) = (IV)}$.

This yields the  total potential as

\eqn{  U_\mr{DCP}\bkt{\mb{r}}=& -\bkt{2n_\mr{th }\bkt{\omega_0}+1}\frac{\mu_0\omega_0^2\bkt{\alpha\bkt{\omega_0 }}^2}{2} \non\\
&\sbkt{\mbc{E}_0\bkt{\mb{r} } \cdot\re{ \dbar{G}_\mr{sc}\bkt{\mb{r} , \mb{r} , \omega_0 }}\cdot \mbc{E}_0^\ast\bkt{\mb{r} }}
}

This is the driven CP potential as given in Eq.\eqref{udcp}, which is in agreement with the result in \cite{Fuchs18} at zero temperature.

We can similarly also find the surface-modified scattering rate as
\eqn{\gamma_\mr{sc} \bkt{\mb{r}}=& \frac{ \mu_0\omega_0^2\bkt{\alpha\bkt{\omega_0 }}^2}{\hbar} \bkt{2n_\mr{th }\bkt{\omega_0}+1} \non\\
&\sbkt{\mbc{E}_0\bkt{\mb{r} } \cdot \im{\dbar{G}_\mr{sc}\bkt{\mb{r} , \mb{r} , \omega_0 }}\cdot \mbc{E}_0^\ast\bkt{\mb{r}}}.
}
For a particle in the near-field limit of a surface with permittivity $\epsilon_S\bkt{\omega}$, for $n_\mr{th}\bkt{\omega_0 }\ll1$ the above can be approximated as
\eqn{\label{gammasc}
\gamma_\mr{sc}\bkt{\mb{r}}\approx\frac{(\alpha\bkt{\omega_0})^2\abs{\mbc{E}_0}^2 }{8\pi\hbar\epsilon_0  \tilde z^3} \im \sbkt{\frac{\epsilon_S\bkt{\omega_0 }-1}{\epsilon_S\bkt{\omega_0 }+1}}.
}

\section{Derivation of the dissipation and noise kernels}
\label{Appdissnoise}
Let us consider the two-time correlation functions of the bath operators as follows,

\begin{widetext}
\eqn{
&\avg{\tilde {\mc{B}} \bkt{t}\tilde {\mc{B}} \bkt{t - \tau}} \non\\
=&  \avg{\sbkt{ {\mb{P}}_0 \bkt{\mb{r}_0, t} \cdot \pard{}{z}\tilde{\mb{E}}_f \bkt{\mb{r}_0, t} +  \pard{}{z}\tilde{\mb{P}}_f \bkt{\mb{r}_0, t} \cdot {\mb{E}}_0 \bkt{\mb{r}_0, t}+\pard{}{z}\mb{P}_0 \bkt{\mb{r}_0, t} \cdot \tilde{\mb{E}}_f \bkt{\mb{r}_0}+ \tilde{\mb{P}}_f \bkt{\mb{r}_0} \cdot\pard{}{z}{\mb{E}}_0 \bkt{\mb{r}_0, t}}\right.\non\\
&\left.\sbkt{ {\mb{P}}_0 \bkt{\mb{r}_0, t-\tau} \cdot \pard{}{z}\tilde{\mb{E}}_f \bkt{\mb{r}_0, t-\tau}+ \pard{}{z}\tilde{\mb{P}}_f \bkt{\mb{r}_0, t-\tau} \cdot {\mb{E}}_0 \bkt{\mb{r}_0, t-\tau}\right.\right.\non\\
&\left.\left.+\pard{}{z}\mb{P}_0 \bkt{\mb{r}_0, t-\tau} \cdot \tilde{\mb{E}}_f \bkt{\mb{r}_0, t-\tau }+ \tilde{\mb{P}}_f \bkt{\mb{r}_0, t-\tau} \cdot\pard{}{z}{\mb{E}}_0 \bkt{\mb{r}_0, t-\tau}} }.
}
\end{widetext}
 We further divide the correlator above into sixteen parts as follows

\eqn{C_1\bkt{\tau } &\equiv \avg{ \sbkt{ \mb{P}_0 \bkt{\mb{r}_0, t} \cdot \pard{}{z}\tilde{\mb{E}}_f \bkt{\mb{r}_0,t} }\right.\non\\
&\left.\sbkt{ \mb{P}_0 \bkt{\mb{r}_0, t-\tau} \cdot \pard{}{z}\tilde{\mb{E}}_f \bkt{\mb{r}_0, t-\tau}}}
}
\eqn{C_2\bkt{\tau } &\equiv \avg{ \sbkt{ \mb{P}_0 \bkt{\mb{r}_0, t} \cdot \pard{}{z}\tilde{\mb{E}}_f \bkt{\mb{r}_0,t} }\right.\non\\
&\left.\sbkt{\pard{}{z}\tilde{\mb{P}}_f \bkt{\mb{r}_0, t-\tau} \cdot \mb{E}_0 \bkt{\mb{r}_0, t-\tau}}}}
\eqn{C_3\bkt{\tau } &\equiv \avg{ \sbkt{\pard{}{z}\tilde{\mb{P}}_f \bkt{\mb{r}_0, t} \cdot \mb{E}_0 \bkt{\mb{r}_0, t}}\right.\non\\
&\left.\sbkt{ \mb{P}_0 \bkt{\mb{r}_0, t-\tau} \cdot \pard{}{z}\tilde{\mb{E}}_f \bkt{\mb{r}_0, t-\tau}}}}

\eqn{
C_4\bkt{\tau } &\equiv\avg{ \sbkt{\pard{}{z}\tilde{\mb{P}}_f \bkt{\mb{r}_0, t} \cdot {\mb{E}}_0 \bkt{\mb{r}_0, t}}\right.\non\\
&\left.\sbkt{ \pard{}{z}\tilde{\mb{P}}_f \bkt{\mb{r}_0, t-\tau} \cdot {\mb{E}}_0 \bkt{\mb{r}_0, t-\tau}}}}

\eqn{C_5\bkt{\tau } &\equiv \avg{ \sbkt{ \pard{}{z}\mb{P}_0 \bkt{\mb{r}_0, t} \cdot \tilde{\mb{E}}_f \bkt{\mb{r}_0,t} }\right.\non\\
&\left.\sbkt{ \pard{}{z} \mb{P}_0 \bkt{\mb{r}_0, t-\tau} \cdot\tilde{\mb{E}}_f \bkt{\mb{r}_0, t-\tau}}}
}
\eqn{C_6\bkt{\tau } &\equiv \avg{ \sbkt{ \pard{}{z}\mb{P}_0 \bkt{\mb{r}_0, t} \cdot \tilde{\mb{E}}_f \bkt{\mb{r}_0,t} }\right.\non\\
&\left.\sbkt{\tilde{\mb{P}}_f \bkt{\mb{r}_0, t-\tau} \cdot \pard{}{z}\mb{E}_0 \bkt{\mb{r}_0, t-\tau}}}}
\eqn{C_7\bkt{\tau } &\equiv \avg{ \sbkt{\tilde{\mb{P}}_f \bkt{\mb{r}_0, t} \cdot \pard{}{z}\mb{E}_0 \bkt{\mb{r}_0, t}}\right.\non\\
&\left.\sbkt{ \pard{}{z}\mb{P}_0 \bkt{\mb{r}_0, t-\tau} \cdot \tilde{\mb{E}}_f \bkt{\mb{r}_0, t-\tau}}}}

\eqn{
C_8\bkt{\tau } &\equiv\avg{ \sbkt{\tilde{\mb{P}}_f \bkt{\mb{r}_0, t} \cdot \pard{}{z}{\mb{E}}_0 \bkt{\mb{r}_0, t}}\right.\non\\
&\left.\sbkt{ \tilde{\mb{P}}_f \bkt{\mb{r}_0, t-\tau} \cdot \pard{}{z}{\mb{E}}_0 \bkt{\mb{r}_0, t-\tau}}}}

\eqn{C_9\bkt{\tau } &\equiv \avg{ \sbkt{ \mb{P}_0 \bkt{\mb{r}_0, t} \cdot \pard{}{z}\tilde{\mb{E}}_f \bkt{\mb{r}_0,t} }\right.\non\\
&\left.\sbkt{ \pard{}{z}\mb{P}_0 \bkt{\mb{r}_0, t-\tau} \cdot \tilde{\mb{E}}_f \bkt{\mb{r}_0, t-\tau}}}
}
\eqn{C_{10}\bkt{\tau } &\equiv \avg{ \sbkt{ \mb{P}_0 \bkt{\mb{r}_0, t} \cdot \pard{}{z}\tilde{\mb{E}}_f \bkt{\mb{r}_0,t} }\right.\non\\
&\left.\sbkt{\tilde{\mb{P}}_f \bkt{\mb{r}_0, t-\tau} \cdot\pard{}{z} \mb{E}_0 \bkt{\mb{r}_0, t-\tau}}}}
\eqn{C_{11}\bkt{\tau } &\equiv \avg{ \sbkt{\pard{}{z}\tilde{\mb{P}}_f \bkt{\mb{r}_0, t} \cdot \mb{E}_0 \bkt{\mb{r}_0, t}}\right.\non\\
&\left.\sbkt{\pard{}{z} \mb{P}_0 \bkt{\mb{r}_0, t-\tau} \cdot \tilde{\mb{E}}_f \bkt{\mb{r}_0, t-\tau}}}}

\eqn{
C_{12}\bkt{\tau } &\equiv\avg{ \sbkt{\pard{}{z}\tilde{\mb{P}}_f \bkt{\mb{r}_0, t} \cdot {\mb{E}}_0 \bkt{\mb{r}_0, t}}\right.\non\\
&\left.\sbkt{ \tilde{\mb{P}}_f \bkt{\mb{r}_0, t-\tau} \cdot\pard{}{z} {\mb{E}}_0 \bkt{\mb{r}_0, t-\tau}}}}

\eqn{C_{13}\bkt{\tau } &\equiv \avg{ \sbkt{\pard{}{z} \mb{P}_0 \bkt{\mb{r}_0, t} \cdot \tilde{\mb{E}}_f \bkt{\mb{r}_0,t} }\right.\non\\
&\left.\sbkt{ \mb{P}_0 \bkt{\mb{r}_0, t-\tau} \cdot \pard{}{z}\tilde{\mb{E}}_f \bkt{\mb{r}_0, t-\tau}}}
}
\eqn{C_{14}\bkt{\tau } &\equiv \avg{ \sbkt{\pard{}{z} \mb{P}_0 \bkt{\mb{r}_0, t} \cdot \tilde{\mb{E}}_f \bkt{\mb{r}_0,t} }\right.\non\\
&\left.\sbkt{\pard{}{z}\tilde{\mb{P}}_f \bkt{\mb{r}_0, t-\tau} \cdot \mb{E}_0 \bkt{\mb{r}_0, t-\tau}}}}
\eqn{C_{15}\bkt{\tau } &\equiv \avg{ \sbkt{\tilde{\mb{P}}_f \bkt{\mb{r}_0, t} \cdot\pard{}{z} \mb{E}_0 \bkt{\mb{r}_0, t}}\right.\non\\
&\left.\sbkt{ \mb{P}_0 \bkt{\mb{r}_0, t-\tau} \cdot \pard{}{z}\tilde{\mb{E}}_f \bkt{\mb{r}_0, t-\tau}}}}

\eqn{
C_{16}\bkt{\tau } &\equiv\avg{ \sbkt{\tilde{\mb{P}}_f \bkt{\mb{r}_0, t} \cdot\pard{}{z} {\mb{E}}_0 \bkt{\mb{r}_0, t}}\right.\non\\
&\left.\sbkt{ \pard{}{z}\tilde{\mb{P}}_f \bkt{\mb{r}_0, t-\tau} \cdot {\mb{E}}_0 \bkt{\mb{r}_0, t-\tau}}}}

such that $\avg{\tilde {\mc{B}} \bkt{t}\tilde {\mc{B}} \bkt{t - \tau}} = \sum_{j = 1}^{16} C_j\bkt{\tau } $.

We now consider the first term closely as follows
\begin{widetext}
\eqn{&C_1\bkt{\tau } = \avg{ \sbkt{ \mb{P}_0 \bkt{\mb{r}_0, t} \cdot \pard{}{z}\tilde{\mb{E}}_f \bkt{\mb{r}_0,t} }\sbkt{ \mb{P}_0 \bkt{\mb{r}_0, t-\tau} \cdot \pard{}{z}\tilde{\mb{E}}_f \bkt{\mb{r}_0, t-\tau}}}\\
=& \frac{1}{4}    \biggl<   \sbkt{\alpha\bkt{\omega_0 }\mbc{E}_0 \bkt{\mb{r}_0} e^{- i \omega_0 t}+ \alpha\bkt{\omega_0 }\mbc{E}_0 ^\ast\bkt{\mb{r}_0} e^{i \omega_0 t}}\cdot\non \\
&\sbkt{\int\dd\omega_1 \sum_{\lambda_1 = e,m} \int\dd^3 r_1 \cbkt{\pard{}{z}\dbar {G}_{\lambda_1} \bkt{ \mb{r}_0,\mb{r}_1, \omega_1}\cdot \hat{\mb{f}}_{\lambda_1}\bkt{\mb{r}_1,  \omega_1}e^{-i \omega_1 t} + \hat{\mb{f}}^\dagger_{\lambda_1}\bkt{\mb{r}_1,  \omega_1}\cdot\pard{}{z}\dbar {G}_{\lambda_1}^\dagger \bkt{ \mb{r}_0,\mb{r}_1, \omega_1} e^{i \omega_1 t}}}\times\nonumber\\
&\sbkt{\int\dd\omega_2 \sum_{\lambda_2 = e,m} \int\dd^3 r_2 \cbkt{\pard{}{z}\dbar {G}_{\lambda_2} \bkt{ \mb{r}_0,\mb{r}_2, \omega_1}\cdot \hat{\mb{f}}_{\lambda_2}\bkt{\mb{r}_2,  \omega_2}e^{-i \omega_2 (t- \tau)} + \hat{\mb{f}}^\dagger_{\lambda_2}\bkt{\mb{r}_2,  \omega_2}\cdot\pard{}{z}\dbar {G}_{\lambda_2}^\dagger \bkt{ \mb{r}_0,\mb{r}_2, \omega_2} e^{i \omega_2 (t- \tau)}}}\cdot \non\\
&\sbkt{\alpha\bkt{\omega_0 }\mbc{E}_0 \bkt{\mb{r}_0} e^{- i \omega_0 (t-\tau)}+ \alpha\bkt{\omega_0 }\mbc{E}_0^\ast \bkt{\mb{r}_0}e^{i \omega_0 (t- \tau)}} \biggr>\\
\label{C1t}
 =& \frac{\mu_0 \hbar}{2\pi}\cos\bkt{\omega_0 \tau }\int\dd\omega\,\omega^2\bkt{\alpha\bkt{\omega_0 }}^2\sbkt{\mbc{E}_0 \bkt{\mb{r}_0}\cdot\partial_z \im \dbar{G}\bkt{\mb{r}_0  , \mb{r}_0 , \omega}\partial_z\cdot \mbc{E}_0^\ast \bkt{\mb{r}_0}} \sbkt{n_\mr{th}\bkt{\omega} e^{i \omega \tau } + \cbkt{n_\mr{th}\bkt{\omega}+ 1} e^{-i \omega \tau } } ,
}
\end{widetext}
where we have averaged over the thermal state of the field and used the fluctuation-dissipation relation for the Green's tensor in obtaining the third step.

Similarly we obtain 
\begin{widetext}
\eqn{\label{C2t}& C_2 \bkt{\tau }  \non\\
&= \frac{\mu_0 \hbar}{2\pi}\cos\bkt{\omega_0 \tau }\int\dd\omega\,\omega^2\alpha\bkt{\omega_0 }\alpha\bkt{\omega }\sbkt{\mbc{E}_0 \bkt{\mb{r}_0}\cdot\partial_z \im \dbar{G}\bkt{\mb{r}_0  , \mb{r}_0 , \omega}\partial_z\cdot \mbc{E}_0^\ast \bkt{\mb{r}_0}} \sbkt{n_\mr{th}\bkt{\omega} e^{i \omega \tau } + \cbkt{n_\mr{th}\bkt{\omega}+ 1} e^{-i \omega \tau } }\\
\label{C3t}
 &C_3 \bkt{\tau }  \non\\
 &= \frac{\mu_0 \hbar}{2\pi}\cos\bkt{\omega_0 \tau }\int\dd\omega\,\omega^2\alpha\bkt{\omega_0 }\alpha\bkt{\omega }\sbkt{\mbc{E}_0 \bkt{\mb{r}_0}\cdot\partial_z \im \dbar{G}\bkt{\mb{r}_0  , \mb{r}_0 , \omega}\partial_z\cdot \mbc{E}_0^\ast \bkt{\mb{r}_0}} \sbkt{n_\mr{th}\bkt{\omega} e^{i \omega \tau } + \cbkt{n_\mr{th}\bkt{\omega}+ 1} e^{-i \omega \tau } }\\
 &C_4 \bkt{\tau }  \non\\
 \label{C4t}
 &= \frac{\mu_0 \hbar}{2\pi}\cos\bkt{\omega_0 \tau }\int\dd\omega\,\omega^2\bkt{\alpha\bkt{\omega }}^2\sbkt{\mbc{E}_0 \bkt{\mb{r}_0}\cdot\partial_z \im \dbar{G}\bkt{\mb{r}_0  , \mb{r}_0 , \omega}\partial_z\cdot \mbc{E}_0^\ast \bkt{\mb{r}_0}} \sbkt{n_\mr{th}\bkt{\omega} e^{i \omega \tau } + \cbkt{n_\mr{th}\bkt{\omega}+ 1} e^{-i \omega \tau } }
 }
 Summing together Eq.\,\eqref{C1t}, \eqref{C2t}, \eqref{C3t}, and \eqref{C4t}, we get
\eqn{\label{C14}\sum_{j = 1}^4C_j\bkt{\tau }=&  \frac{\mu_0 \hbar }{2\pi} \cos\bkt{\omega_0 \tau } \int \dd\omega\,\omega^2\sbkt{\alpha\bkt{\omega_0 } + \alpha\bkt\omega}^2  \sbkt{\mbc{E}_0 \bkt{\mb{r}_0}\cdot\cbkt{ \partial_z \im\dbar{G}  \bkt{ \mb{r}_0 , \mb{r}_0 , \omega}\partial_z}\cdot \mbc{E}_0^\ast \bkt{\mb{r}_0}}\non\\
&\sbkt{\cbkt{n_\mr{th}\bkt{\omega}+1}e^{-i \omega\tau} + n_\mr{th}\bkt{\omega} e^{i \omega\tau} }.
} 
Similarly, it can be shown that 
\eqn{\label{C58}\sum_{j = 5}^8C_j\bkt{\tau }=&  \frac{\mu_0 \hbar }{2\pi} \cos\bkt{\omega_0 \tau } \int \dd\omega\,\omega^2\sbkt{\alpha\bkt{\omega_0 } + \alpha\bkt\omega}^2  \sbkt{\partial_z\mbc{E}_0 \bkt{\mb{r}_0}\cdot  \im\dbar{G}  \bkt{ \mb{r}_0 , \mb{r}_0 , \omega}\cdot \partial_z\mbc{E}_0^\ast \bkt{\mb{r}_0}}\non\\
&\sbkt{\cbkt{n_\mr{th}\bkt{\omega}+1}e^{-i \omega\tau} + n_\mr{th}\bkt{\omega} e^{i \omega\tau} }\\
\label{C912}
\sum_{j = 9}^{12}C_j\bkt{\tau }=& \frac{\mu_0 \hbar }{2\pi} \cos\bkt{\omega_0 \tau } \int \dd\omega\,\omega^2\sbkt{\alpha\bkt{\omega_0 } + \alpha\bkt\omega}^2  \sbkt{\mbc{E}_0 \bkt{\mb{r}_0}\cdot\cbkt{  \partial_z\im\dbar{G}  \bkt{ \mb{r}_0 , \mb{r}_0 , \omega}}\cdot \partial_z\mbc{E}_0^\ast \bkt{\mb{r}_0}}\non\\
&\sbkt{\cbkt{n_\mr{th}\bkt{\omega}+1}e^{-i \omega\tau} + n_\mr{th}\bkt{\omega} e^{i \omega\tau} }\\
\label{C1316}
\sum_{j = 13}^{16}C_j\bkt{\tau }=&   \frac{\mu_0 \hbar }{2\pi} \cos\bkt{\omega_0 \tau } \int \dd\omega\,\omega^2\sbkt{\alpha\bkt{\omega_0 } + \alpha\bkt\omega}^2  \sbkt{\partial_z\mbc{E}_0 \bkt{\mb{r}_0}\cdot  \cbkt{\im\dbar{G}  \bkt{ \mb{r}_0 , \mb{r}_0 , \omega}\partial_z}\cdot \mbc{E}_0^\ast \bkt{\mb{r}_0}}\non\\
&\sbkt{\cbkt{n_\mr{th}\bkt{\omega}+1}e^{-i \omega\tau} + n_\mr{th}\bkt{\omega} e^{i \omega\tau} }
}

Eq.\,\eqref{C14}, \eqref{C58}, \eqref{C912}, and \eqref{C1316} yield
\eqn{\label{BBt}\avg{ \tilde{\mc{B}}\bkt{t-\tau}\tilde{\mc{B}}\bkt{t}}
=& \frac{\mu_0\hbar }{2\pi } \cos\bkt{\omega_0 \tau } \int \dd\omega\,\omega^2\sbkt{\alpha\bkt{\omega_0 } + \alpha\bkt\omega}^2 \sbkt{\cbkt{n_\mr{th}\bkt{\omega}+1}e^{-i \omega\tau} + n_\mr{th}\bkt{\omega} e^{i \omega\tau} } g\bkt{\mb{r}_0 , \omega},
}
where we have defined

\eqn{
g\bkt{\mb{r}_0 , \omega}\equiv &\sbkt{  \mbc{E}_0 \bkt{\mb{r}_0}\cdot\cbkt{\partial_z \im\dbar{G}  \bkt{ \mb{r}_0 , \mb{r}_0 , \omega}\partial_z}\cdot \mbc{E}_0^\ast \bkt{\mb{r}_0}+ \partial_z \mbc{E}_0 \bkt{\mb{r}_0}\cdot\im\dbar{G}  \bkt{ \mb{r}_0 , \mb{r}_0 , \omega}\cdot \partial_z\mbc{E}_0^\ast \bkt{\mb{r}_0}\right.\non\\
&\left.\mbc{E}_0 \bkt{\mb{r}_0}\cdot\cbkt{\partial_z \im\dbar{G}  \bkt{ \mb{r}_0 , \mb{r}_0 , \omega}}\cdot \partial_z\mbc{E}_0^\ast \bkt{\mb{r}_0}+ \partial_z \mbc{E}_0 \bkt{\mb{r}_0}\cdot\cbkt{\im\dbar{G}  \bkt{ \mb{r}_0 , \mb{r}_0 , \omega} \partial_z}\cdot\mbc{E}_0^\ast \bkt{\mb{r}_0}}.
}

Similarly it can be shown

\eqn{\label{BtB}\avg{ \tilde{\mc{B}}\bkt{t-\tau}\tilde{\mc{B}}\bkt{t}}
=& \frac{\mu_0\hbar }{2\pi } \cos\bkt{\omega_0 \tau } \int \dd\omega\,\omega^2\sbkt{\alpha\bkt{\omega_0 } + \alpha\bkt\omega}^2 \sbkt{\cbkt{n_\mr{th}\bkt{\omega}+1}e^{i \omega\tau} + n_\mr{th}\bkt{\omega} e^{-i \omega\tau} }g\bkt{\mb{r}_0 , \omega}
}

Using \eqref{BBt} and \eqref{BtB} we can write the dissipation and noise kernels as

\eqn{
\mc{D}(\tau ) =& i\sbkt{\avg{ \tilde {\mc{B}}\bkt{t}\tilde {\mc{B}}\bkt{t-\tau}}-\avg{ \tilde {\mc{B}}\bkt{t-\tau}\tilde {\mc{B}} \bkt{t}}}\non\\ =&\frac{\mu_0 \hbar}{\pi } \cos\bkt{\omega_0 \tau } \int \dd\omega\,\omega^2\sbkt{\alpha\bkt{\omega_0 } + \alpha\bkt\omega}^2\sin\bkt{\omega \tau}g\bkt{\mb{r}_0 , \omega},\\
\mc{N}(\tau ) =&\avg{ \tilde {\mc{B}}\bkt{t}\tilde {\mc{B}}\bkt{t-\tau}}+\avg{ \tilde {\mc{B}}\bkt{t-\tau}\tilde {\mc{B}} \bkt{t}}\non\\
=&\frac{\mu_0 \hbar}{\pi } \cos\bkt{\omega_0 \tau } \int \dd\omega\,\omega^2\sbkt{\alpha\bkt{\omega_0 } + \alpha\bkt\omega}^2\sbkt{2n_{\mr{th}}\bkt{\omega } + 1}\cos\bkt{\omega \tau}g\bkt{\mb{r}_0 , \omega}.
}
\end{widetext}
Further noting that $2n_\mr{th}\bkt{\omega} +1 = \coth\bkt{\frac{\hbar \omega}{2k_B T}}$, we arrive at Eq.\,\eqref{diss} and \eqref{noise}, with the spectral density given by \eqref{Jw}.
\section{Decoherence from  other sources}
\label{Appdecother}
As a reference, we compare the surface-fluctuation induced decoherence with that due to other sources that can potentially be a limiting mechanism for preparing macroscopic quantum states as follows.
\subsection{Background gas scattering}
The decoherence rate  due to background gas scattering \cite{MaxBook} is given as
\eqn{\Lambda_\mr{ gas}= \frac{8}{3\hbar^2 } P_\mr{gas} \bkt{2 \pi m_\mr{gas} }^{1/2} R^2 \bkt{k_B T}^{1/2},
} 
where $P_\mr{gas}$ is the gas pressure, and $m_\mr{gas}\approx 5\times 10^{-26} $\,kg is the mass of a single gas molecule. For a background gas pressure of  $P_\mr{gas} \sim 1$\,mbar--$10^{-11}$\,mbar, one obtains  a corresponding localization parameter $\Lambda_\mr{gas}\sim 10^{33}$\,$\mr{Hz}/\text{m}^2$--$10^{20}$\,$\mr{Hz}/\text{m}^2$. One potential way to circumvent decoherence due to background gas scattering could be to perform the experiment on time scales shorter than those of average successive collisions of the system with gas molecules \cite{Skatepark}.

\subsection{Scattering of blackbody radiation}
The COM decoherence of the dielectric nanosphere induced due to scattering of blackbody radiation is given as \cite{MaxBook}
\eqn{ \Lambda_\mr{BB} =\frac{8 ! c}{18 \pi} \sbkt{\frac{\alpha\bkt{\omega_\mr{th}}}{\pi \epsilon_0 }}^2\bkt{\frac{k_B T}{\hbar c}}^9 \zeta \bkt{9},
}
where $\zeta \bkt{9}\approx1.002$ refers to the Riemann $\zeta$-function, $\omega_\mr{th}\equiv \frac{2 \pi c T}{b}$ is the peak blackbody radiation frequency, with $b$ as the Wien's displacement constant. For $T \sim  1$\,K--100\,K, we find the blackbody radiation induced decoherence as $\Lambda_\mr{BB} \sim 10^{-7}\,\mr{Hz}/\mr{m}^2$--$10^{11}\,\mr{Hz}/\mr{m}^2$. We note that this corresponds to the decoherence arising from a purely thermal background, and for large enough temperatures can potentially exceed the decoherence from the driven CP interaction as we can see from Fig.\,\ref{Fig2}\,(b).


\vspace{-0.2cm}\footnotesize{

}
\end{document}